\documentclass[10pt,twocolumn,superscriptaddress,english,10pt,prb,showpacs,floatfix]{revtex4-1}
\usepackage{amsthm}
\usepackage{amsmath}
\usepackage{amssymb}
\usepackage{mathdots}
\usepackage{graphicx}

\bibpunct{[}{]}{,}{n}{}{}

\makeatletter

\@ifundefined{textcolor}{}
{%
 \definecolor{BLACK}{gray}{0}
 \definecolor{WHITE}{gray}{1}
 \definecolor{RED}{rgb}{1,0,0}
 \definecolor{GREEN}{rgb}{0,1,0}
 \definecolor{BLUE}{rgb}{0,0,1}
 \definecolor{CYAN}{cmyk}{1,0,0,0}
 \definecolor{MAGENTA}{cmyk}{0,1,0,0}
 \definecolor{YELLOW}{cmyk}{0,0,1,0}
}

\usepackage{braket}
\usepackage[backref=true,
bookmarksnumbered=true,
bookmarks=true,
bookmarksopen=true,
colorlinks=true,
citecolor=blue,
linkcolor=blue,
anchorcolor=green,
urlcolor=blue,unicode=false]{hyperref}

\newcommand{\abs}[1]{\left| #1 \right|} 
 
\let\baraccent=\= 
\renewcommand{\=}[1]{\stackrel{#1}{=}} 

\DeclareMathOperator\Sgn{Sgn}

\makeatother

\usepackage{babel}

\begin{document}
\title{Signatures of topological Josephson junctions}

\author{Yang Peng}
\affiliation{\mbox{Dahlem Center for Complex Quantum Systems and Fachbereich Physik, Freie Universit{\"a}t Berlin, 14195 Berlin, Germany} }

\author{Falko Pientka}
\affiliation{\mbox{Dahlem Center for Complex Quantum Systems and Fachbereich Physik, Freie Universit{\"a}t Berlin, 14195 Berlin, Germany} }
\affiliation{Department of Physics, Harvard University, Cambridge MA 02138, USA}

\author{Erez Berg}
\affiliation{Department of Condensed Matter Physics, Weizmann Institute of Science, Rehovot, Israel}

\author{Yuval Oreg}
\affiliation{Department of Condensed Matter Physics, Weizmann Institute of Science, Rehovot, Israel}

\author{Felix von Oppen}
\affiliation{\mbox{Dahlem Center for Complex Quantum Systems and Fachbereich Physik, Freie Universit{\"a}t Berlin, 14195 Berlin, Germany} }

\begin{abstract}
Quasiparticle poisoning and diabatic transitions may significantly narrow the window for the experimental observation of the $4\pi$-periodic $dc$ Josephson effect predicted for topological Josephson junctions. Here, we show that switching current measurements provide accessible and robust signatures for topological superconductivity which persist in the presence of quasiparticle poisoning processes. Such measurements provide access to the phase-dependent subgap spectrum and Josephson currents of the topological junction when incorporating it into an asymmetric SQUID together with a conventional Josephson junction with large critical current. We also argue that pump-probe experiments with multiple current pulses can be used to measure the quasiparticle poisoning rates of the topological junction. The proposed signatures are particularly robust, even in the presence of Zeeman fields and  spin-orbit coupling, when focusing on short Josephson junctions.  Finally, we also consider microwave excitations of short topological  Josephson junctions which may complement switching current measurements. 
\end{abstract}

\maketitle

\section{Introduction}

Topological superconductors with $p$-wave pairing and Majorana bound states \cite{kitaev01} are currently attracting much interest, motivated in part by
possible applications to topological quantum information processing \cite{kitaev03}. Several solid-state platforms have been proposed  \cite{fu08,fu09,lutchyn10,oreg10,alicea11,beenakker_magnetic,flensberg,morpurgo,bernevig13,pientka13} and are vigorously pursued experimentally \cite{mourik12,das12,churchill13,rokhinson12,lund,harlingen,yacoby,kouwenhoven,ali,ruby,wiedenmann2016,paajaste2016}. A key question of current research is to develop appropriate detection schemes which allow one to identify topological superconducting phases and Majorana bound states. 

A particularly striking signature of topological superconductivity is provided by Josephson junctions 
formed by a weak link between two topological superconductors hosting unpaired Majorana bound states at their ends \cite{kitaev01,fu09}. While for conventional superconductors, the Josephson current is $2\pi$ periodic in the applied phase difference, the Josephson current across a junction made from topological superconductors is predicted to be $4\pi$ periodic \cite{kitaev01}. This period doubling of the Josephson current in a topological Josephson junction \cite{fu08,fu09,badiane11,liang2011, aguado2012,pikulin12,beenakker2013,aguado2013,zhang2014,michaeli2014,braggio2015} is protected by fermion number parity and as such quite sensitive to quasiparticle poisoning which changes the occupation of subgap states by inelastic processes involving the quasiparticle continuum. If the temporal variation of the superconducting phase difference across the junction is too slow, quasiparticle poisoning restores the $2\pi$ periodicity \cite{fu09}. If the phase difference is varied too fast, the periodicity is restored by diabatic transitions into the quasiparticle continuum \cite{badiane11}.  

Here, we explore an alternative approach to probe the phase-dependent subgap spectrum of a topological Josephson junction, which is inspired by a recent series of remarkable experiments on conventional Josephson junctions \cite{Rocca07,zgirski11,bretheau13,bretheau13nature}. These experiments consider Josephson junctions based on atomic weak links which host localized subgap Andreev states.  The experiments explore the phase-dependent subgap spectrum by switching-current measurements as well as microwave spectroscopy. Here, we establish that analogous experiments provide a promising technique to distinguish between conventional and topological Josephson junctions. We find that this is particularly true in the short-junction limit, i.e., for junctions which are short compared to the coherence length of the adjacent (topological) superconductors. An important advantage of such measurements is that they can be performed in the presence of quasiparticle poisoning and in fact explicitly exploit processes that break fermion parity. 

Ideally, Josephson junctions carry a dissipationless supercurrent (or Josephson current) as long as the applied current remains below the critical current and switch to a resistive state once the current exceeds the critical current \cite{tinkham}. In practice, the switching current fluctuates about the critical current due to thermal fluctuations. This has characteristic consequences in switching-current measurements based on applying short current pulses. Indeed, the switching probability as a function of the height of the applied current pulse does not increase abruptly from zero to one at the critical current, but rather exhibits a smooth step when accounting for fluctuations arising from the electromagnetic environment. When the junction hosts subgap states, their occupations also fluctuate due to quasiparticle poisoning processes. The current-phase relation and hence the critical current depend on the occupation of the subgap states, so that poisoning processes lead to fluctuations in the switching current.

The effect of poisoning processes is particularly simple when the current pulses are short compared to typical poisoning processes. In this case, the poisoning dynamics determines the occupation probability of the various subgap states prior to applying the current pulse but does not modify the state during the pulse duration. The switching probability becomes a superposition of step functions corresponding to the various subgap occupations. When the broadening of the steps is small compared to the shifts in the switching current between different occupation states, the measured switching probability exhibits a sequence of steps -- one for each occupation of the subgap states -- with intermediate plateaus. The heights of the plateaus reflect the occupation probabilities of the various junction states at the beginning of the current pulse. As a consequence, the switching probability encodes information on the current-phase relations for the various occupation states of the Josephson junction. 

\begin{figure}[t]
    \includegraphics[width=0.35\textwidth]{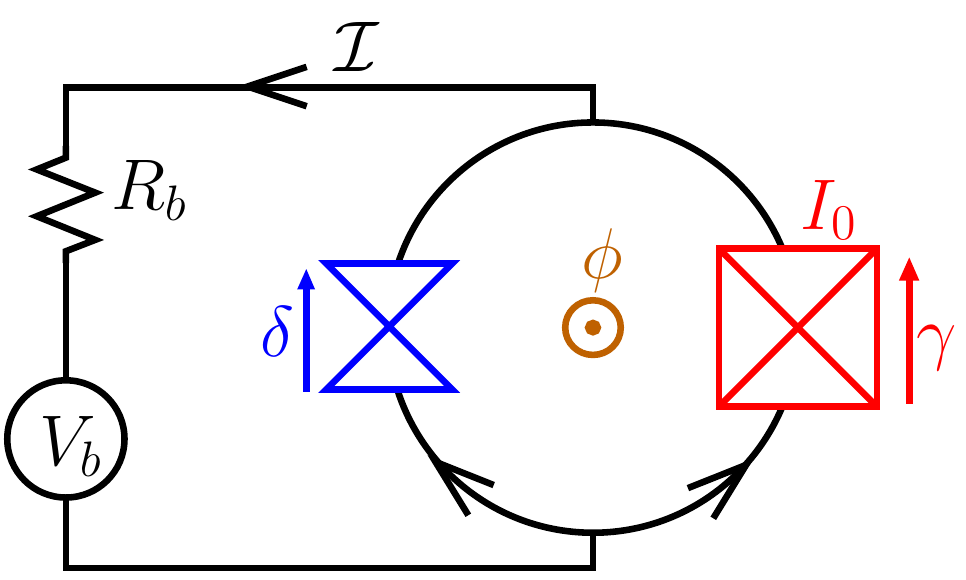} 
    \caption{\label{fig.SQUID} Basic setup of the asymmetric SQUID, involving a weak conventional/topological Josephson junction
    (blue triangles) and a strong auxiliary Josephson junction (red checked box) with critical current $I_0$. The phase $\delta$ across the weak junction is linked to the phase
    $\gamma$ across the auxiliary junction and the phase drop $\phi=2e\varphi/\hbar$ induced by the magnetic flux $\varphi$ threading the SQUID loop, $\delta =  \phi + \gamma $. The applied voltage $V_b$ drives a current $\mathcal{I}$ through the resistance $R_b$ and the
  SQUID.} 
\end{figure}

Switching-current measurements on a single Josephson junction do not provide access to the phase dependence of the Josephson current, but merely to the maximal Josephson current and its dependence on the junction occupation. Phase-dependent information can be obtained by incorporating the junction of interest into an asymmetric SQUID where the second auxiliary junction in the SQUID loop has a much larger critical current and no subgap states \cite{Rocca07}. This setup is illustrated in Fig.\ \ref{fig.SQUID}. The switching current of the SQUID is shifted away from the switching current of the large junction by the phase-dependent Josephson current of the weak one, so that switching-current measurements as a function of flux can provide access to the entire current-phase relation of the various states of the weak junction of interest. 

This makes switching-current measurements suitable to probe a unique distinction between topological and conventional Josephson junctions. As a function of the phase difference $\delta$ across the junction, 
the difference in Josephson currents between different junction states must vanish an even number of times within a $2\pi$ period in a trivial junction, and an odd number of times in a topological junction. 

In this scheme, the initial occupation probability of the various junction states is assumed thermal. When driving the system out of equilibrium, switching-current measurements also provide access to the poisoning dynamics \cite{zgirski11}. Imagine that the system is taken out of equilibrium at some initial time $t=0$ so that the occupation probability of the various subgap states is no longer thermal. Poisoning processes will subsequently induce relaxation to equilibrium, and the rate of this relaxation can be probed by switching-current measurements after a time delay $t$. This pump-probe scheme can either be implemented by a sequence of two current pulses with time delay $t$, or by applying an appropriate microwave pulse at time $t=0$ prior to the switching current measurements at time $t$. 

Microwave irradiation also provides an alternative spectroscopic way of measuring the subgap spectrum as it induces transitions between different occupation states of the Josephson junction by microwave radiation \cite{bretheau13nature,kos13,virtanen2013,bretheau2014,vayrynen2015}. Thus, evidence for topological superconductivity can be further strengthened by performing switching-current measurements in conjunction with microwave spectroscopy. This motivates us to calculate the admittance of a topological Josephson junction in the short-junction limit, complementing the results of Ref.\ \cite{vayrynen2015} for the long-junction limit. 

Such measurements provide various opportunities to distinguish topological from nontopological Josephson junctions. We find that the signatures are particularly distinctive for short junctions as their subgap spectrum contains only few Andreev states. Such short topological junctions support only a single subgap state at energy $E_M$ (and its particle-hole conjugate at $-E_M$), originating from the hybridization of the two Majorana bound states. In contrast, a short conventional junction frequently (but not necessarily) supports additional Andreev states associated with the spin degree of freedom. In this case, topological and nontopological junctions can be distinguished by the number of plateaus in the switching probability as a function of applied current. Only junctions with a single plateau are suspects for being topological [see Fig.\ \ref{figwidth}(a)]. Among these suspects, the subgap spectrum exhibits a fermion-parity protected level crossing at a phase difference of $\delta=\pi$ for topological junctions, and an anticrossing for nontopological junctions. Thus, the Josephson current at a phase difference of $\pi$ is maximal for topological junctions and vanishes  for conventional ones. This leads to characteristic differences in the flux dependence of the plateau width [see Fig.\ \ref{figwidth}(b)]. Finally, even if the anticrossing of a nontopological junction happens to be too weak to be resolved, its poisoning dynamics should be characteristically different. Poisoning dynamics necessarily involves the quasiparticle continuum for topological junctions while poisoning processes involving only subgap states can exist for conventional junctions. These signatures based on switching current measurements can be further corroborated by microwave spectroscopy.

The paper is organized as follows. In Sec.~\ref{Sec:Heuristic} we review basic considerations on the differences between the topological and conventional Josephson junctions. Sec.~\ref{Sec:Psw} contains the central results of this paper. After introducing  the asymmetric SQUID setup we discuss the characteristic distinctions between topological and conventional Josephson junctions in switching-current measurements, including the effects of thermal fluctuations in the context of the RCSJ model. We end this section with a discussion of pump-probe experiments with multiple current pulses which provide access to the quasiparticle poisoning rates. Microwave absorption is discussed for short topological junctions based on 2D topological insulators in Sec.~\ref{Sec:Theory}. While we discuss nontopological junctions in the absence of Zeeman fields or spin-orbit coupling in the earlier sections, these couplings are typically present in experiments searching for possible topological superconductivity. We show in Sec.~\ref{Sec:TJJ_spin} that the signatures distinguishing topological from nontopological junctions remain robust in the presence of these effects when focusing on the short-junction limit. Finally, we conclude in Sec.~\ref{Sec:con}.

\section{Basic considerations 
\label{Sec:Heuristic}}

\subsection{Conventional Josephson junction}

\begin{figure}[t]
    \includegraphics[width=0.22\textwidth]{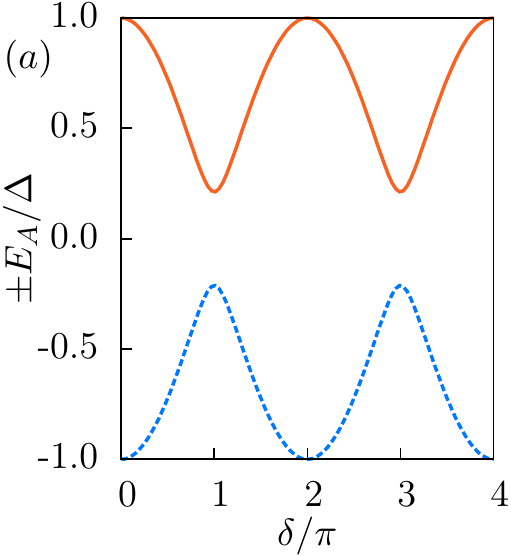}
    \includegraphics[width=0.22\textwidth]{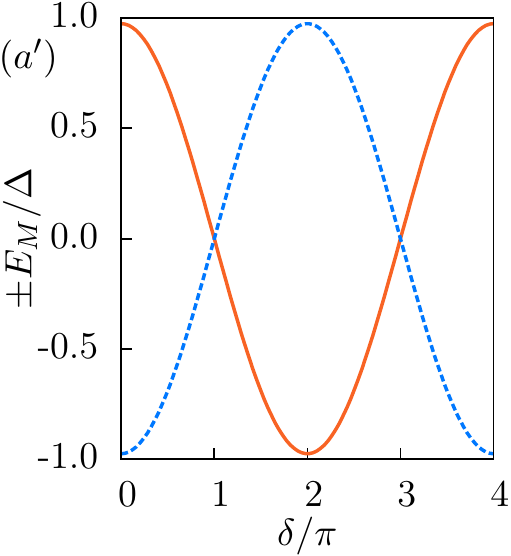}
    \includegraphics[width=0.22\textwidth]{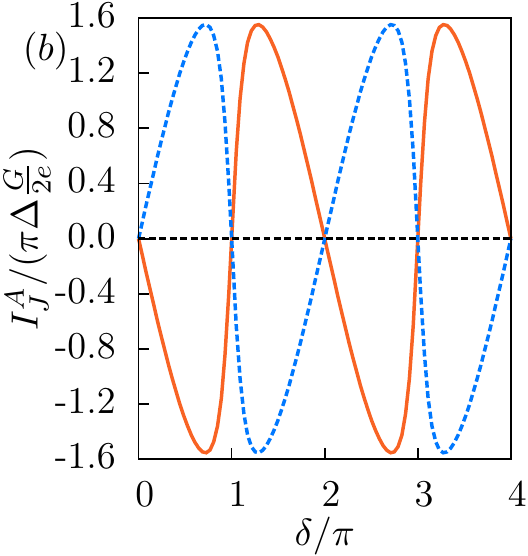}
    \includegraphics[width=0.22\textwidth]{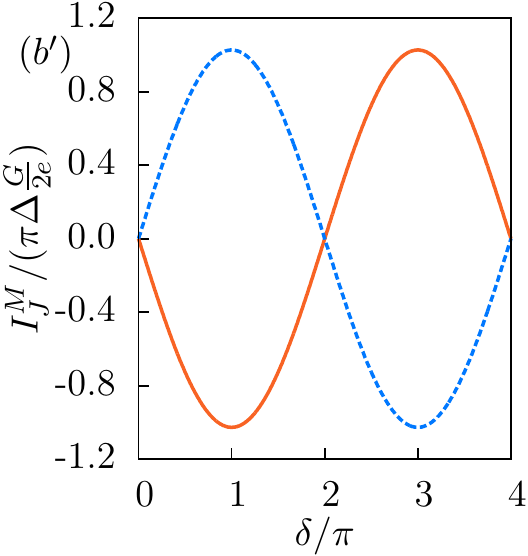}
    \caption{\label{fig.comparison} Upper panels:  Single-particle energies of the subgap state as a function of the phase difference across the junction for (a) conventional and
    (a$^\prime$) topological Josephson junctions. Lower panels: Supercurrent as a function of phase difference for the various possible states of (b) a conventional and (b$^\prime$)
    a topological Josephson junctions ($G=e^2D/\pi$, $D=0.95$). The blue, black, and red curves in (b) display the currents for the states $\ket{0}$, $\ket{1,\sigma}$, and
    $\ket{2}$, respectively. The blue and red curves in (b$^\prime$) display the currents for the states $\ket{0}$ and $\ket{1}$. }
\end{figure}

To set the stage, we first review the case of a conventional Josephson junction. As realized in experiment \cite{Rocca07,zgirski11,bretheau13}, we consider a short junction (i.e., shorter than the superconducting coherence length) in the single-channel limit. If this channel has transmission $D$, the junction binds a single, spin-degenerate Andreev bound state at subgap energy \cite{beenakker91}
\begin{equation}
E_A(\delta) = \Delta\sqrt{1-D\sin^2\frac{\delta}{2}}.
\label{eq:ABS_energy}
\end{equation}
Here, $\delta$ denotes the phase difference across the junction and $\Delta$ the superconducting gap. Figure \ref{fig.comparison}(a) shows this particle-hole symmetric pair $E = \pm E_A(\delta)$ of Bogoliubov-de Gennes states as a function of the phase difference $\delta$. 

In the absence of above-gap excitations, these single-particle subgap states give rise to four many-body states associated with the Josephson junction. In the ground state, denoted by $\ket{0}$, the positive-energy Andreev bound state is empty. In addition, there are two degenerate excited states in which either the spin-up or the spin-down Andreev level is occupied. We denote these states by $\ket{1\uparrow}=\gamma_\uparrow^\dagger\ket{0}$ and $\ket{1\downarrow}=\gamma_\downarrow^\dagger\ket{0}$, where $\gamma_{\uparrow}$ and $\gamma_{\downarrow}$ are the Bogoliubov operators associated with the Andreev state. Finally, the Andreev state can be doubly occupied, $\ket{2}=\gamma^\dagger_\uparrow\gamma^\dagger_\downarrow\ket{0}$. Note that the states $\ket{0}$ and $\ket{2}$ are even states in terms of fermion parity, while $\ket{1\uparrow}$ and $\ket{1\downarrow}$ are odd. 

In equilibrium, the Josephson current is governed by the many-body energy ${\cal E}(\delta)$ of the
junction. In the ground state $\ket{0}$, the (phase-dependent) junction energy is given by $-E_A(\delta)$. Correspondingly, the two odd states $\ket{1\uparrow}$ and $\ket{1\downarrow}$ have zero energy, while the doubly-occupied state $\ket{2}$ has energy $+E_A(\delta)$. This is summarized as 
\begin{equation}
    {\cal E}_n(\delta) = (n-1) E_A(\delta),
\end{equation}
where $n=0,1,2$ denotes the occupancy of the Andreev bound state. The Josephson current in state $\ket{n}$ can be obtained from the energy as
\begin{equation}
    {\cal I}_n(\delta)=  2e \frac{\partial {\cal E}_n(\delta)}{\partial\delta}= 2(n-1)e\frac{\partial  E_A(\delta)}{\partial\delta}.
    \label{eq:In_conventional}
\end{equation}
Thus, the Josephson currents of the two states $\ket{0}$ and $\ket{2}$ have the same magnitude, but flow in opposite directions, while the Josephson current vanishes in the odd states $\ket{1\uparrow}$ and $\ket{1\downarrow}$. The $2\pi$-periodic supercurrents for these states are shown in Fig.~\ref{fig.comparison}(b). 

\subsection{Topological Josephson junction}

The corresponding results for topological Josephson junctions differ in several essential ways. Here, we focus attention on junctions made of topological superconductors which 
break time reversal symmetry and are hence characterized by unpaired Majorana bound states at their ends. The simplest
realization of such a topological superconducting phase occurs in spinless $p$-wave superconductors
\cite{kitaev01,review1, review2}. These phases can for instance be realized experimentally based on two-dimensional topological insulators proximity coupled to conventional $s$-wave superconductors \cite{fu09} or semiconductor quantum wires \cite{lutchyn10,oreg10}. When tuned to the right parameter regime, these systems realize phases which are  adiabatically connected to the topological phase of spinless $p$-wave superconductors and are thus promising venues for realizing the topological Josephson junction setup which we propose. Indeed, several experiments have already investigated such Josephson junctions with the goal of identifying signatures of topological superconductivity \cite{rokhinson12,yacoby,kouwenhoven,wiedenmann2016}.  

In the following, we assume that any ungapped normal part of the junction region is short compared to the coherence length $\xi$ of the adjacent topological superconducting phase. Then, the subgap spectrum emerges from two overlapping Majorana bound states localized at the ends of the two topological superconductors \cite{fu09,lutchyn10,oreg10,alicea11}. This yields one nondegenerate Andreev level $E_M(\delta)$. While $E_M(\delta)$ is $4\pi$ periodic, the overall particle-hole-symmetric subgap spectrum $\pm E_M(\delta)$ is $2\pi$ periodic. Moreover, the level crossings between $E_M(\delta)$ and $-E_M(\delta)$ at $\delta$ equal to odd multiples of $\pi$ are protected by conservation of fermion parity. This single-particle spectrum is shown in Fig.~\ref{fig.comparison}(a$^\prime$).

As the topological Josephson junction has a single nondegenerate Andreev state, there are only two rather than four many-body states in the absence of above-gap quasiparticle excitations. We denote the state in which the Andreev level $E_M(\delta)$ is empty (occupied) as $\ket{0}$ ($\ket{1}$). The two states satisfy $\ket{1}=\gamma^\dagger\ket{0}$, where $\gamma$ is the Bogoliubov operator associated with the subgap state $E_M(\delta)$. We will also refer to $\ket{0}$ as having even fermion parity or the even state and to $\ket{1}$ as the odd state. (In the presence of above-gap quasiparticles, both occupations are however accessible for any parity of the electron number. Such processes are known as quasiparticle poisoning.) 

The phase-dependent many-body energy of the junction is equal to $-E_M(\delta)/2$ for the even state $\ket{0}$ and $+E_M(\delta)/2$ for the odd state 
$\ket{1}$, or
\begin{equation}
    {\cal E}_n(\delta)= (2n-1) \frac{E_M(\delta)}{2} 
\end{equation}
for state $\ket{n}$ with $n=0,1$ denoting the occupation of the Andreev state. Just as the Bogoliubov-de Gennes states, the two many-body states $\ket{n}$ become degenerate for $\delta$ equal to odd multiples of $\pi$. Notice that the odd state can have lower energy than the even state as $E_M(\delta)$ becomes negative, which cannot happen in a conventional Josephson junction.

The Josephson current follows from the many-body energy in the usual way, so that
\begin{equation}
    {\cal I}_n(\delta) =  2e \frac{\partial {\cal E}_n(\delta)}{\partial \delta} =  e (2n-1) \frac{\partial E_M(\delta)}{\partial \delta}.
    \label{eq:In_topological}
\end{equation}
For fixed fermion parity $n$, the Josephson current is $4\pi$ periodic, as illustrated in Fig.~\ref{fig.comparison}(b$^\prime$). The two states carry supercurrents of the same magnitude but of opposite sign. 

This implies that there are distinct differences in the supercurrent carried by conventional and topological Josephson junctions. Unlike a conventional Josephson junction, a topological Josephson junction does not have states with zero Josephson current. Moreover, conventional Josephson junctions can assume three different current states, while topological junctions are limited to two states. We will explore experimental consequences of these differences in Sec.\ \ref{Sec:Psw}.

\subsection{Excitation spectra}

\begin{figure}[t]
 \includegraphics[height=0.23\textwidth]{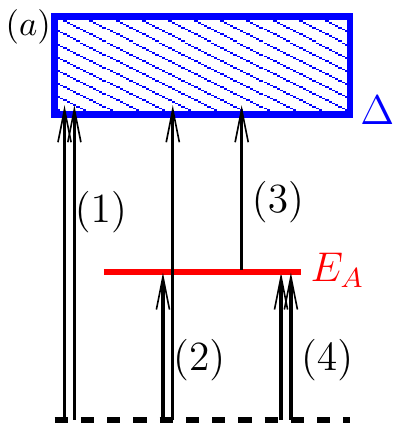}
 \includegraphics[height=0.23\textwidth]{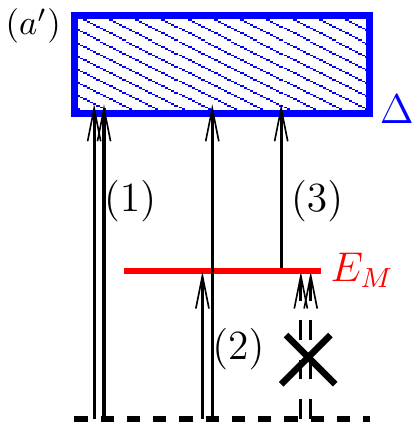}
 \includegraphics[height=0.23\textwidth]{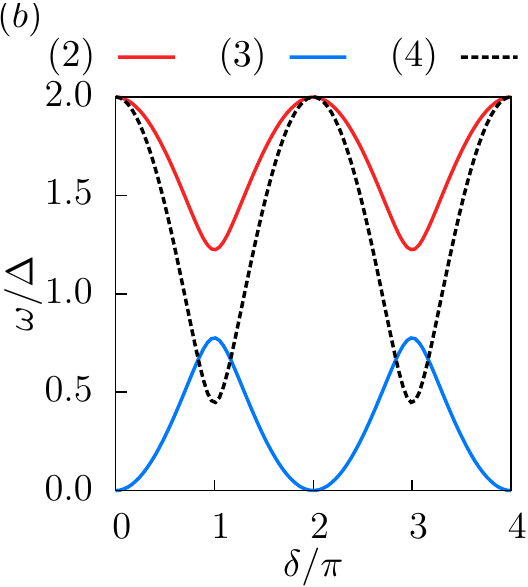}
 \includegraphics[height=0.23\textwidth]{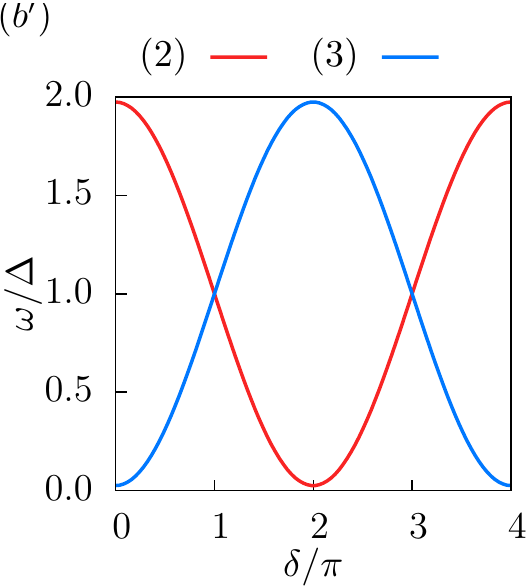}
 \caption{\label{fig.process} Upper panels: Possible quasiparticle processes numbered by (1)-(4) in 
(a) conventional and (a$^\prime$) topological Josephson junctions. The black dashed lines indicate the many-body ground state
and the upper blue boxes the quasiparticle continuum above the energy gap $\Delta$. The red lines indicate the bound state at energies $E_A$ or $E_M$ for conventional and
topological junctions, respectively. Lower panels: Excitation energies (or energy thresholds) involving the bound state corresponding to the various processes in panels (a) and
(a$^\prime$). }
\end{figure}

The differences in subgap structures are also reflected in the excitation spectrum of the junction under microwave irradiation. Continuing to focus on short junctions, the many-body energy of a conventional junction can assume three different values. Correspondingly, the subgap states lead to three resonances in the differential absorption of microwave irradiation, as shown in Fig.~\ref{fig.process}(a) \cite{kos13}. In the absence of subgap states, the only excitation process that breaks up a Cooper pair excites both electrons into the quasiparticle continuum, see process (1) in Fig.\ \ref{fig.process}, which has a threshold energy of $2\Delta$. The existence of subgap states allows for the following additional processes. In process (2), a Cooper pair in the condensate is split, with one of the quasiparticles excited into the bound state at energy $E_A$ and the second into the quasiparticle continuum above the gap $\Delta$. This process has threshold energy $E_A+\Delta$. Process (3) corresponds to a quasiparticle in the bound state being excited into the continuum. This process has threshold energy $\Delta-E_A$. Finally, process (4) splits a Cooper pair, with both quasiparticles getting excited into the bound state. This process requires a threshold energy of $2E_A$. The thresholds of processes (2)-(4) are sketched in Fig.\ \ref{fig.process}(b) as a function of the phase $\delta$ across the junction. We note in passing that these considerations are valid for zero magnetic field. The more general case will be considered in Sec.\ \ref{Sec:TJJ_spin}.

A topological Josephson junction allows fewer microwave-induced transitions involving subgap states as it can only assume two possible junction energies \cite{vayrynen2015}. When the junction is in the even-parity state, a Cooper pair can be split, with one electron occupying the subgap level and the second getting excited into the quasiparticle continuum. This process requires a threshold photon energy of $\Delta+E_M$ and is labeled as process (2) in Fig.\ \ref{fig.process}(a$^\prime$). When the junction is in the odd-parity state, the quasiparticle occupying the Andreev state $E_M(\delta)$ can be excited to the quasiparticle continuum. This process, labeled as (3) in Fig.\ \ref{fig.process}(a$^\prime$), requires a threshold energy of $\Delta-E_M$. While these two processes are similar to corresponding processes in conventional Josephson junctions, there is no analog of process (4). Indeed, there is only a single, nondegenerate Andreev level in topological Josephson junctions and it is impossible to split a Cooper pair exciting both electrons into a subgap state. A more complete theory of the microwave absorption is presented in Sec.\ \ref{Sec:Theory}.  

\section{Switching probability of topological Josephson junctions\label{Sec:Psw}}

In this section, we explore the consequences of the qualitative differences between the subgap spectra of conventional and topological superconductors for switching-current measurements of asymmetric SQUIDs. We first present a heuristic approach in Sec.\ \ref{sec:heuristic}. As illustrated in Fig.\ \ref{fig.Psw1}, we find that there are characteristic differences between short topological and conventional junctions both in the number and the width of the plateaus in the switching probability. These schematic results are further corroborated by detailed numerical results in Sec.\ \ref{sec:sign}, based on the RCSJ theory developed in Sec.\ \ref{RCSJ}, with the central results shown in Fig.\ \ref{fig.Psw2}. Finally, in Sec.\ \ref{sec:pois}, we propose pump-probe approaches to the switching probability to explore the poisoning dynamics and show that this encodes further characteristic differences between topological and nontopological Josephson junctions. 

\subsection{Plateaus in the switching probability}
\label{sec:heuristic}

Consider the SQUID device shown in Fig.~\ref{fig.SQUID}, consisting of a large auxiliary Josephson junction and the weak junction of interest which can be either conventional or topological. The auxiliary Josephson junction is assumed to have a large critical current $I_0$ and no internal dynamics. The weak junction of interest has a much smaller critical current and internal dynamics associated with the bound-state occupation, as discussed in the previous section. The phase differences across the large junction (denoted by $\gamma$) and the weak junction (denoted by $\delta$) are related through \begin{equation}
  \delta=\phi+\gamma,
  \label{PhaseRelation}
\end{equation} 
where $\phi=2e\varphi/\hbar$ is the phase drop induced by the magnetic flux $\varphi$ threading the SQUID loop. (This relation assumes that the geometric inductance of the SQUID loop can be neglected as in recent experiments \cite{zgirski11}.)

The total applied current ${\cal I}$ flowing through the SQUID splits between the auxiliary junction with current  
\begin{equation}
    {\cal I}_{\rm aux}(\gamma)=I_0\sin \gamma,
\end{equation}
and the weak junction of interest with current ${\cal I}_{n}(\delta)$,
\begin{equation}
    {\cal I}={\cal I}_{\rm aux}(\gamma)+{\cal I}_{n}(\phi+\gamma).
    \label{eq:current_conservation}
\end{equation}
Here, we have used the relation (\ref{PhaseRelation}) between the phase differences. For zero applied current, ${\cal I}=0$, the current circulates around the SQUID loop and both junctions carry the same current, albeit with opposite signs. As the auxiliary junction has a much larger critical current, its phase difference $\gamma$ is small and the phase drop $\phi$ due to the flux is applied almost entirely to the weak junction, i.e., $\delta \simeq \phi$. 

When a current bias ${\cal I}$ is applied to the junction, the auxiliary junction carries most of this current and we can first focus on its behavior. Then, the phase difference across this junction is approximately 
\begin{equation}
  \gamma \simeq \arcsin \frac{{\cal I}}{I_0},
\end{equation} 
and the junction becomes resistive when the current exceeds the critical current ${\cal I}^{\rm aux}_{\rm sw}=I_0$ of the junction. Ideally, this occurs when $\gamma$ reaches $\gamma_{\rm sw}=\pi/2$. 

\begin{figure}[t]
    \includegraphics[width=0.45\textwidth]{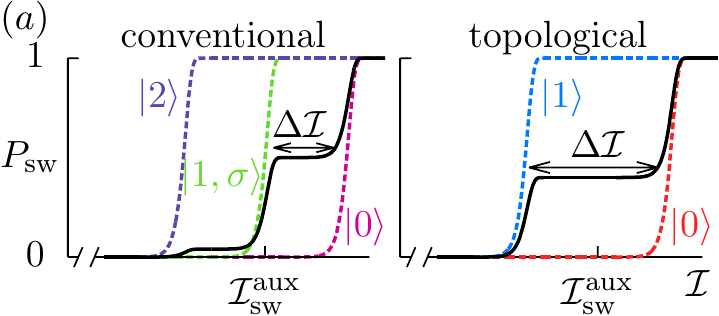}
    \includegraphics[width=0.45\textwidth]{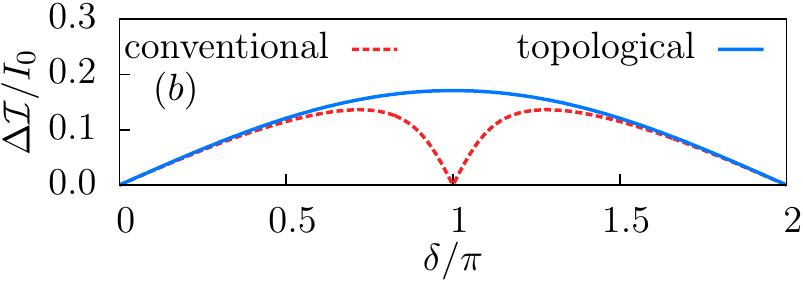}
    \caption{\label{fig.Psw1} (a) Probability $P_{\rm sw}$ of switching to the resistive state as a function of current for conventional (left) and topological (right) Josephson
    junctions for $\delta=0.9\pi$ and $D=0.95$. The dashed lines are the switching probabilities for the junction assuming a fixed occupation state, cf.\ Eq.~(\ref{eq:Psw_state}).
    The black solid curves display the switching probability $P_{\rm sw}$ in the presence of quasiparticle poisoning, and can be obtained from a weighted average over the switching
    probabilities of the various occupation states, cf.\ Eq.~(\ref{eq:Psw_aver}). For the conventional Josephson junction, we choose the weight factors $c_{0}=0.5$,
    $c_{1,\uparrow}=c_{1,\downarrow}=0.23$, and $c_{2}=0.04$. For the topological Josephson junction, we choose the weight factors $c_{0}=0.6$ and $c_{1}=0.4$. (b) Width of the plateau $\Delta {\cal I}/I_0$ as a function of $\delta=\gamma+\phi$ for the case of conventional  (red dashed) and topological Josephson junctions (blue solid) (for $E_J^{\rm aux}/\Delta=5.7$, where $E_J^{\rm aux}=\hbar I_0/2e$ is the Josephson energy of the auxiliary junction and $\Delta$ the gap of the weak junction).}
\label{figwidth}
\end{figure}

In the presence of the weak junction, switching occurs for the value of $\gamma=\gamma_{\rm sw}$ for which the right hand side of Eq.\ (\ref{eq:current_conservation}) has its maximum. Expanding to linear order in the small current ${\cal I}_n$, we have
\begin{equation}
  \gamma_{\rm sw} \simeq \frac{\pi}{2} + \frac{1}{I_0}\left. \frac{{\mathrm d}{\cal I}_n(\phi+\gamma)}{{\mathrm d}\gamma}\right|_{\gamma=\pi/2}
\end{equation}
and 
\begin{equation}
    {\cal I}_{\rm sw} \simeq {\cal I}_0+{\cal I}_{n}(\phi+\pi/2)
\end{equation}
for the switching current. This relation implies that the switching current of the SQUID reflects the current-phase relation of the weak junction. A measurement of the switching current of the asymmetric SQUID as a function of flux $\phi$ can therefore be used to measure this current-phase relation. 

As shown in Ref.\ \cite{zgirski11} for a nontopological Josephson junction based on an atomic contact, this can be used to resolve the current-phase relation of the various many-body states of the junction.
Indeed, if the switching-current measurement is performed sufficiently fast compared to quasiparticle poisoning processes in the weak junction, the switching current of the SQUID depends on the occupancy $n$ of the Andreev state. In practice \cite{Rocca07,zgirski11,bretheau13,bretheau13nature}, this measurement can be performed by applying short current pulses and measuring the probability that the SQUID switches into the resistive state as a function of applied current ${\cal I}$. In the simple approximation given here, this probability has the step-like form
\begin{equation}
    P^n_{\rm sw}({\mathcal I},\phi)= \theta({\cal I} - I_0 - {\cal I}_n(\phi+\pi/2))
\label{eq:Psw_state}
\end{equation}
when the weak junction is in state $n$. The switching probabilities -- for fixed $n$ and $\phi$ -- are illustrated by the dashed lines in Fig.~\ref{fig.Psw1}(a), which includes plots for both conventional and topological Josephson junctions. (The steps in the switching probability as a function of current are smoothed in this figure, anticipating the more elaborate model discussed in Sec.\ \ref{RCSJ}.) The critical current ${\cal I}^{\rm aux}_{\rm sw}=I_0$ of the auxiliary junction by itself is marked on the $x$-axis. According to Eq.\ (\ref{eq:Psw_state}), the shifts of the steps away from this value can be interpreted as the supercurrent flowing through the weak junction of interest. 

The dashed lines in Fig.~\ref{fig.Psw1}(a) assume that the junction of interest is in a specific state $n$ in the beginning of the current pulse (and that this charge state does not get modified during the pulse duration). In practice, the state of the junction changes statistically due to quasiparticle poisoning processes and is in general uncontrolled in experiment. Thus, the junction has probability $c_n$ to be in state $n$ at the beginning of the current pulse. If we keep assuming that the junction does not switch between states over the duration of the current pulse, the experimentally measured switching probability  
\begin{equation}
    P_{\rm sw}(\mathcal{I},\phi)=\sum_{n}c_n P^n_{\rm sw}({\cal I},\phi)
    \label{eq:Psw_aver}
\end{equation}
is a weighted average over the states $n$ of the junction. Such weighted averages are illustrated by black full lines in Fig.~\ref{fig.Psw1}(a).

In the simplest approximation, the probabilities $c_n$ can be assumed as thermal. More generally, they can be obtained from rate equations which describe the relevant poisoning processes \cite{olivares2014,zazunov2014}. Remarkably, one does not need detailed information about this poisoning kinetics for establishing robust signatures of topological superconductivity. Indeed, as illustrated in Fig.~\ref{fig.Psw1}(a), the weighted average exhibits plateaus as a function of current. The number of plateaus increases with the number of current states of the junction. A conventional Josephson junction can have three different current states, and will then exhibit {\em two} plateaus in a plot of the switching probability versus current. In contrast, a short topological junction has only two current states and thus merely a {\em single} plateau. Thus, if all junction states are occupied with an appreciable probability $c_n$, topological and nontopological junctions frequently differ in the number of plateaus. 

However, the number of plateaus may also be the same for topological and nontopological junctions. This happens when one of the $c_n$  is so small (presumably for the $\ket{2}$ state) for a conventional junction that only a single plateau can be resolved, or because the nontopological junction also has only a single subgap state, as can be the case in the presence of Zeeman splitting (see Sec.\ \ref{Sec:TJJ_spin} below for explicit model calculations). Even in this case, however, there remains a clear-cut difference between topological and conventional junctions when considering the width of the plateau as a function of the flux applied to the SQUID. The width of the plateau measures the difference in the supercurrents between the two contributing junction states. 

At the flux $\phi$ such that the phase across the weak junction $\delta$ is equal to $\pi$, the difference in supercurrent is maximal for a topological junction, but vanishes for conventional junctions. Correspondingly, the width of the plateau should be maximal near $\delta=\pi$ for a topological junction, but vanishes for a conventional junction. This central result of this paper is illustrated in Fig.~\ref{fig.Psw1}(b).
Note that the experimental control parameter is $\phi$ rather than $\delta$. However, these are simply related by
$\delta = \phi + \gamma_{\rm sw} \simeq \phi + \pi/2$ at the position of the steps. It is useful
to mention that the plateau width in the topological case is linear in the transmission amplitude $\sqrt{D}$. Thus, the
lower the transmission, the narrower the plateau, making it more difficult to detect and characterize it experimentally.

In the more detailed considerations presented in Sec.\ \ref{sec:sign}, we show that the height and the location of the plateau provide additional criteria for distinguishing topological and conventional junctions.

\subsection{RCSJ model}
\label{RCSJ}

A more accurate description of the asymmetric SQUID is provided by the RCSJ model
\cite{McCumber1968,Stewart1968,Ambegaokar,Kurkijarvi1972,fulton1974,tinkham}, which takes into account its shunting resistance $R_S$ and capacitance $C$. Starting from this model and assuming that the weak junction remains in a particular state $n$, current conservation and the Josephson relation imply that the dynamics of the phase $\gamma$ across the auxiliary junction is described by 
\begin{equation}
    \frac{\hbar C}{2e}\ddot{\gamma} = {\cal I} - I_0 \sin\gamma - \mathcal{I}_{n}(\phi+\gamma) - \frac{\hbar}{2eR_S}\dot{\gamma} + \tilde{\zeta}(t).
\label{eq:RCSJ}
\end{equation}
The term $\tilde{\zeta}(t)$ accounts for the thermal fluctuations associated with the resistance $R_S$ and satisfies 
$\braket{\tilde{\zeta}(t)\tilde{\zeta}(t')}=({2T}/{R_S})\delta(t-t')$ at temperature $T$. Note that Eq.~(\ref{eq:RCSJ}) reduces to Eq.~(\ref{eq:current_conservation}), when neglecting the thermal fluctuations and searching for a solution with time-independent $\gamma$. It is convenient to introduce new parameters through
\begin{gather*}
    m=\left(\frac{\hbar}{2e}\right)^{2}C,\quad
    \eta=\frac{1}{R_SC},\quad\zeta(t)=\frac{\hbar}{2e}\tilde{\zeta}(t)
\end{gather*}
as well as the effective potential
\begin{equation}
    U(\gamma)= - E_J^{\rm aux}\cos \gamma + \mathcal{E}_{n}(\phi+\gamma) - \frac{\hbar \mathcal{I}\gamma}{2e},
\end{equation}    
where $E_J^{\rm aux}=\hbar I_0 /2e$. Then, the equation for the phase $\gamma$ takes the form of a Langevin equation 
\begin{equation}
    m\ddot{\gamma}=-U'(\gamma)-m\eta\dot{\gamma}+\zeta(t)
\end{equation}
for a ``particle" moving in the ``tilted washboard" potential $U(\gamma)$ with friction coefficient $\eta$ and the correlator
\begin{equation}
 \braket{\zeta(t)\zeta(t')}=2Tm\eta\delta(t-t')
\end{equation}
of the Langevin force.

\begin{figure}[t]
    \includegraphics[width=0.35\textwidth]{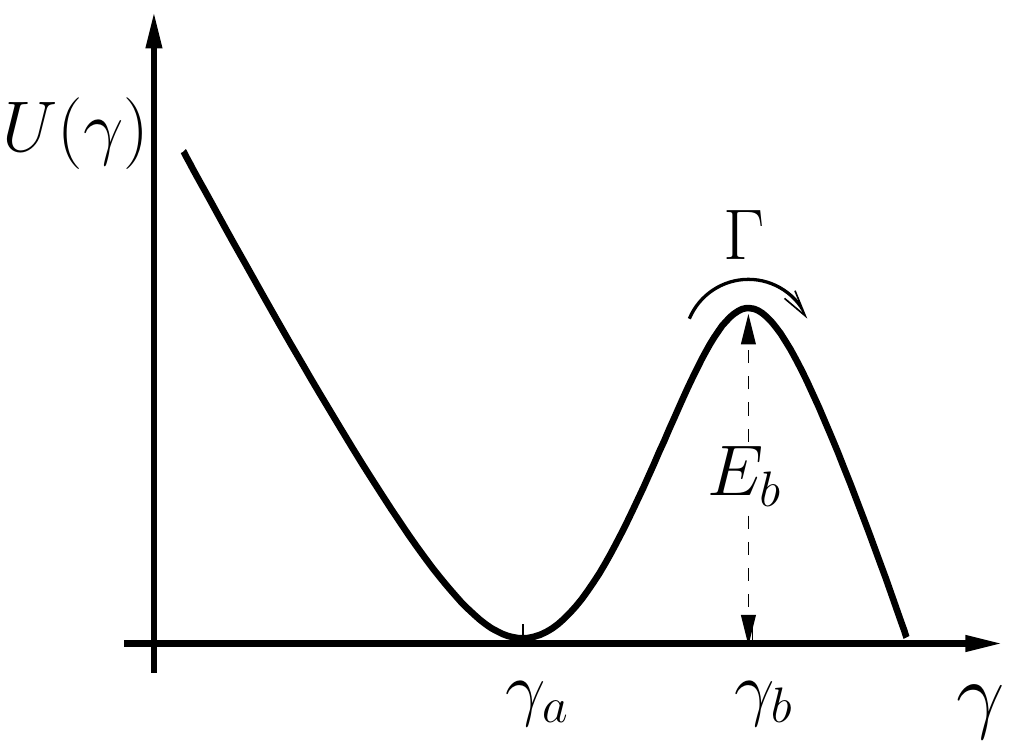}
    \caption{\label{fig.potential} Sketch of the ``tilted washboard'' potential governing the dynamics of the Josephson junction near one minimum.}
\end{figure}

At zero bias current, $\mathcal{I}=0$, $U(\gamma) \simeq -E_J^{\rm aux}\cos\gamma$ and the ``particle" will most likely remain near the potential minimum $\gamma\simeq 0$ (modulo 2$\pi$). With increasing bias current, the potential $U(\gamma)$ is tilted and the particle eventually escapes from the minimum (see Fig.~\ref{fig.potential}), with the SQUID developing a voltage according to the Josephson relation $V=\hbar \dot{\gamma}/2e$. 

The probability $P_{\rm sw}$ that a current pulse of duration $t_p$ switches the junction to a finite-voltage state can be expressed in terms of the escape rate $\Gamma$ from the minimum  \cite{Rocca07,zgirski11,bretheau13},
\begin{equation}
  P_{\rm sw}=1-\exp(-\Gamma t_p).
  \label{eq:Psw_Gamma}
\end{equation}
To determine $\Gamma$, we consider the overdamped limit of the Langevin equation,  
\begin{equation}
\dot{\gamma}=-\left(m\eta\right)^{-1}U'(\gamma)+\left(m\eta\right)^{-1}\zeta(t).
\end{equation}
In this limit, the probability density $\mathcal{P}(\gamma,t)$ of the auxiliary junction's phase difference $\gamma$ is governed by the Smoluchowski equation \cite{Hanggi}
\begin{equation}
    \frac{\partial\mathcal{P}(\gamma,t)}{\partial
    t}=\frac{1}{m\eta}\frac{\partial}{\partial\gamma}\left[U'(\gamma)\mathcal{P}(\gamma,t)+T\frac{\partial\mathcal{P}(\gamma,t)}{\partial\gamma}\right]
\end{equation}
and the escape rate can be computed by Kramers rate theory \cite{Hanggi,Kramers}. 

Consider the minimum of $U(\gamma)$ at $\gamma_{a}$ and the neighboring maximum at $\gamma_b$ (see Fig.\ \ref{fig.potential}). The rate $\Gamma$ can be obtained by solving the stationary Smoluchowski equation at a constant probability current 
\begin{equation}
   j =  \frac{1}{m\eta}\left[U'(\gamma)\mathcal{P}(\gamma,t)+T\frac{\partial\mathcal{P}(\gamma,t)}{\partial\gamma}\right]
\label{jsmol}
\end{equation}
with absorbing boundary condition at $\gamma=\gamma_{+}$, $\mathcal{P}(\gamma_{+})=0$. The position $\gamma_+$ has to be sufficiently far to the right of $\gamma_b$, i.e.,  $\gamma_+>\gamma_b$, but is otherwise arbitrary. Then, the probability current $j$, normalized to the occupation $n_a$ of the minimum at $\gamma_a$, describes the rate $\Gamma$ at which transitions occur out of the minimum $\gamma_a$.

Note that the  Smoluchowski equation implies that $j$ is independent of $\gamma$ for stationary solutions, so that we find 
\begin{equation}
    \mathcal{P}(\gamma)=\frac{m\eta j}{T}\exp\left(-U(\gamma)/T\right)\int_{\gamma}^{\gamma_{+}}dy\,\exp\left(U(y)/T\right) 
\end{equation}
by solving Eq.\ (\ref{jsmol}). For $\gamma$ near $\gamma_a$, we can perform the integral by saddle-point integration and obtain
\begin{equation}
     \mathcal{P}(\gamma)\simeq \frac{m\eta j}{\omega_{b}T}\sqrt{\frac{2\pi T}{m}}\exp\left(\frac{U(\gamma_{b})}{T}\right)\exp\left(-U(\gamma)/T\right).
\label{neargammaa}
\end{equation}
Here, we approximate $U(\gamma)\simeq U(\gamma_{b})-\frac{1}{2}m\omega_{b}^2(\gamma-\gamma_b)^2$ around $\gamma_b$. The population $n_a$ in the potential well around $\gamma_a$  is 
\begin{equation}
    n_a \simeq \int_{-\infty}^{\infty}d\gamma'\,\mathcal{P}(\gamma') = \frac{2\pi\eta j}{\omega_{a}\omega_{b}}\exp\left(E_{b}/T\right),
\end{equation}
where the integral should be evaluated with the expression in Eq.\ (\ref{neargammaa}) We used the expansion $U(\gamma)\simeq U(\gamma_{a})+\frac{1}{2}m\omega_{a}^2(\gamma-\gamma_a)^2$ for $\gamma$ near $\gamma_a$ and introduced the barrier height $E_b=U(\gamma_b)-U(\gamma_a)$. Finally, one obtains the Arrhenius-like expression
\begin{equation}
    \Gamma=\frac{j}{n_{a}}=\frac{\omega_{a}\omega_{b}}{2\pi\eta}\exp\left(-E_{b}/T\right).
    \label{eq:escape_rate}
\end{equation}
for the escape rate $\Gamma$.

\begin{figure}[t]
    \includegraphics[width=0.45\textwidth]{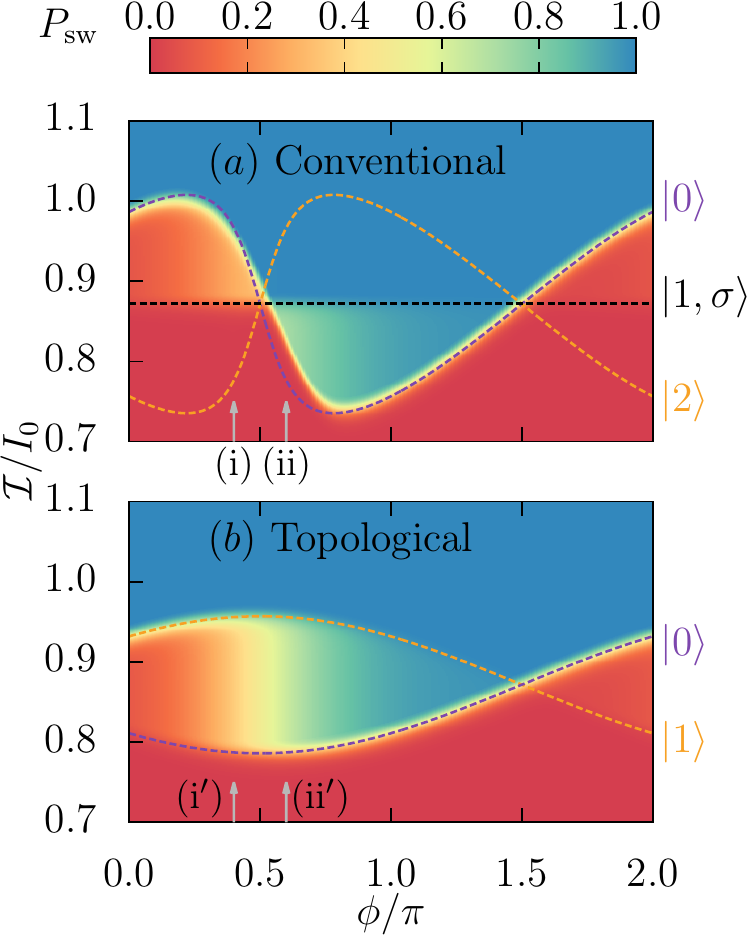}
    \caption{\label{fig.Psw2} Color plot of the switching probability $P_{\rm sw}$ of asymmetric SQUIDs as a function of flux $\phi$ and height ${\cal I}$ of the current pulse for (a) a conventional and (b) a topological Josephson junction. The occupation probabilities of the various junction states prior to the current pulse are taken to be thermal, with effective temperature $T_{\rm eff}$. In (a), parameters are such that the occupation probability of the doubly occupied Andreev state is negligible. The dashed lines indicate the switching currents based on the Josephson currents associated with the various junction states as indicated in the figure, with the phase difference across the weak junction taken as $\delta=\phi+\pi/2$. In (a), the purple line corresponds to the ground state, the black line to the singly-occupied Andreev state, and the orange one to the doubly occupied state. In (b) the purple and orange lines correspond to the two states of the topological junction. The parameters were chosen as $R_{s}=550\Omega$, $I_{0}=553.7{\rm nA}$, $T=100{\rm mK}$, $E_{\rm J}^{\rm aux}/\Delta=5.7$, $t_p = 1\mathrm{\mu s}$ and $D=0.95$, according to the parameters used in Ref.~\cite{zgirski11}. The effective temperature $T_{\rm eff}$ is chosen as such that $E_{J}^{\rm aux}/T_{\rm eff} = 10$.} The grey arrows with labels (i), (ii) indicate values of $\phi$ for which linecuts are shown in Fig.\ \ref{fig.fixphi}.
\end{figure}

The two points $\gamma_{a}$ and $\gamma_{b}$ satisfy the condition $\partial U(\gamma)/\partial\gamma=0$, which yields
\begin{equation}
    I_0\sin\gamma+\mathcal{I}_{n}(\gamma+\phi)=\mathcal{I}.
\label{gammaagammab}
\end{equation}
First neglecting the contribution of the weak junction, one has
\begin{equation}
  \gamma_a \simeq \arcsin \frac{{\cal I}}{I_0} \,\,\, ; \,\,\,
 \gamma_b \simeq \pi - \arcsin \frac{{\cal I}}{I_0}
\label{solgam}
\end{equation}
as well as
\begin{equation}
    \omega_{a}\omega_{b}\simeq \frac{E_{J}^{\rm aux}}{m} |\cos\gamma_a\cos\gamma_b|^{1/2}
        \label{eq:omega_ab} 
\end{equation}
and 
\begin{equation}
      E_{b}\simeq  E_{J}^{\rm aux} (\cos\gamma_{a}-\cos\gamma_{b}) - \frac{\hbar {\cal I}}{2e}(\gamma_b-\gamma_a).
\end{equation}
Then, Eq.\ (\ref{eq:escape_rate})  yields the phase escape rate
\begin{align}
    &\Gamma^{\rm aux}({\cal I})=\frac{eI_0R_S}{\pi\hbar}\sqrt{1-(\mathcal{I}/I_0)^{2}} \nonumber\\
    &\times e^{-\frac{\hbar} {2eT} [ {\cal I} (2\arcsin(\mathcal{I}/I_0)-\pi) + 2I_0 \sqrt{1-(\mathcal{I}/I_0)^{2}}]}
\end{align}
by Eq.\ (\ref{eq:Psw_Gamma}), the switching probability of the auxiliary junction is
\begin{equation}
  P_{\rm sw}^{\rm aux}({\cal I}) = 1- e^{-\Gamma^{\rm aux}({\cal I})t_p}
\end{equation}
$P_{\rm sw}^{\rm aux}$ has a step-like shape as shown in Fig.~\ref{fig.Psw1}, with the steps occurring near $\mathcal{I}_{\rm sw}^{\rm aux}$ which is generally smaller than $I_0$ due to the thermal fluctuations.

\begin{figure}[t]
    \includegraphics[width=0.45\textwidth]{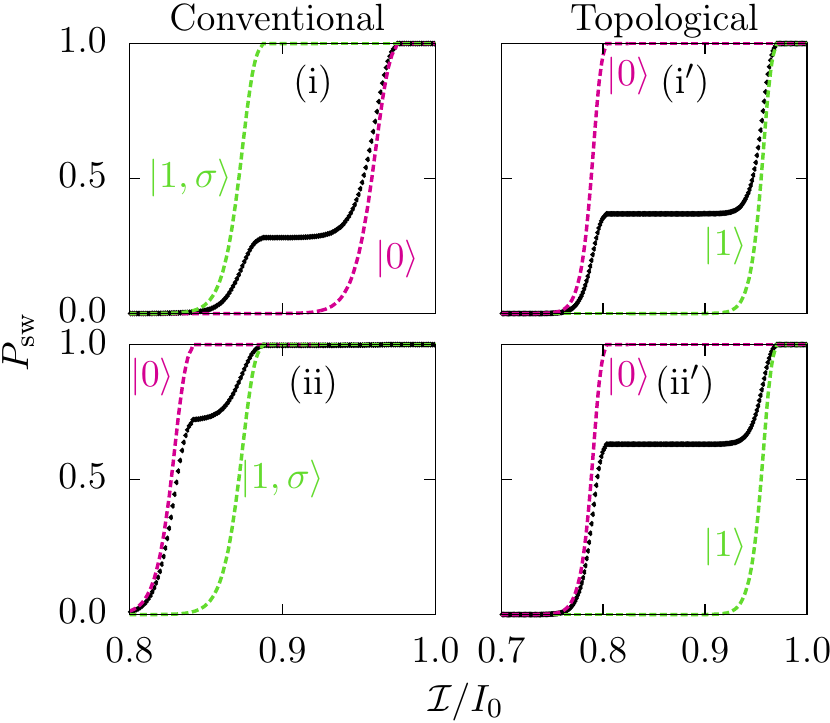}
    \caption{\label{fig.fixphi}Switching probability of a conventional (topological) junction as a function of the applied current for fixed $\phi$. The black symbols represent $P_{\rm sw}$ along the fixed-$\phi$ cuts indicated by grey arrows in Fig.~\ref{fig.Psw2} for conventional junctions: (i) Switching probability for $\phi = 0.4\pi$; (ii) for $\phi = 0.6\pi$. (i$^\prime$) and (ii$^\prime$) show the corresponding plots along the same $\phi$ cuts for the topological junction. The dashed curves denote the switching probability when the weak junction is in the fixed occupation state as specified in the figure, similar to the ones in Fig.~\ref{fig.Psw1}. Note that for conventional junction, the state with the lower switching current inverts between (i) and (ii). It is this inversion which explains the sudden change in the plateau height for (i) $\phi<\pi/2$ and (ii) $\phi>\pi/2$, as discussed in the text. In contrast, there are no such inversions in the topological case.}
\end{figure}

Now, the weak junction can be readily included to first order. We first need to solve Eq.\ (\ref{gammaagammab}) for $\gamma_a$ and $\gamma_b$. In doing so, we can replace $\gamma$ in the argument of ${\cal I}_n$ by the results in Eq.\ (\ref{solgam}) for $\gamma_a$ and $\gamma_b$ to zeroth order. At sufficiently low temperatures, the junction switches only once the barrier becomes small and hence when $\gamma_a$ and $\gamma_b$ are close together (and thus close to $\pi/2$). In computing the switching probability to first order in ${\cal I}_n$, it is sufficient to set $\gamma\simeq \pi/2$ in the argument of ${\cal I}_n$ in Eq.\ (\ref{gammaagammab}). Then, we can account for the weak junction simply by shifting ${\cal I} \to {\cal I} - {\cal I}_n(\phi+\pi/2)$ in the above considerations. This yields 
\begin{equation}
\label{switchres}
  P^n_{\rm sw}({\cal I}) \simeq P_{\rm sw}^{\rm aux}({\cal I}-{\cal I}_n(\phi+\frac{\pi}{2})) 
\end{equation}
for the switching probability of the asymmetric SQUID. 

\subsection{Signatures of topological Josephson junctions}
\label{sec:sign}

The differences between topological and trivial junctions are most pronounced in the switching probablity $P_{\rm sw}$ as a function of the flux and the height of the current pulse. We can use the RCSJ approach developed in the previous section to calculate $P_{\rm sw}$ in Eq.\ (\ref{eq:Psw_aver}) numerically, see Eq.\ (\ref{switchres}). This leads to Fig.\ \ref{fig.Psw2} which contains a central result of this paper and highlights the qualitative difference between topological and  trivial junctions. Panel (a) of Fig.\ \ref{fig.Psw2} shows a color plot of the switching probability for a nontopological junction as a function of the height ${\cal I}$ of the current pulse and the flux threading the SQUID.  The dashed lines indicate the switching currents for the various junction states as obtained on the basis of the current-phase relation of the weak junction and discussed in Sec.\ \ref{sec:heuristic}. The purple line corresponds to the ground state of the junction, the black line to the odd states, and the orange line to the doubly-occupied Andreev level. 

For the parameters chosen, double occupation of the Andreev level can be neglected so that the switching probability effectively exhibits only a single plateau as a function of current. In Fig.\ \ref{fig.Psw2}, this plateau is well resolved for $0\lesssim \phi \lesssim \pi$, corresponding to a phase difference of $\pi/2 \lesssim \delta \lesssim 3\pi/2$ across the weak junction. Outside this region, the energy of the odd states becomes too high -- and their thermal occupation too low -- so that the corresponding step in the switching probability is no longer resolved. Obviously, the range over which the plateau can be resolved depends on the junction parameters and temperature. 

The height of the intermediate plateau changes quite abruptly at $\phi\simeq\pi/2$, corresponding to a phase difference of $\delta=\pi$ across the weak junction. This is seen in Fig.\ \ref{fig.Psw2} and further illustrated in the line cuts presented in Fig.\ \ref{fig.fixphi}. At $\delta=\pi$, there is a change in sign of the Josephson current flowing through the weak junction. Consequently, the low-current step in the switching probability is due to the odd states (ground state) to the left (right) of $\phi=\pi/2$, and the step heights therefore controlled by the low (high) thermal occupations of these states. Note that this change in the plateau height occurs at a flux where the width of the plateau goes through zero.

Corresponding results for a topological junction are shown in panel (b) of Fig.\ \ref{fig.Psw2}. The two dashed lines correspond to the expected switching currents based on the even and odd states of the topological junction. The plateau in the switching probability occurs between these two lines. Unlike for the conventional junction, the width of the plateau is now maximal for $\phi=\pi/2$, corresponding to a phase difference of $\delta=\pi$ across the topological junction. This qualitative difference between topological and conventional junctions was already highlighted in Fig.\ \ref{figwidth}. Note also that there is now a rather abrupt change in the height of the plateau at this point of maximal plateau width, rather than the point of minimal plateau width as for conventional junctions. 

Finally, there are characteristic differences between conventional and topological junctions based on the flux dependence of the Josephson current. In a conventional junction, one of the steps of the switching probability as a function of current is due to the odd state which carries zero Josephson current for all phase differences. Thus, the position of one of the steps should be rather insensitive to the flux $\phi$. Conversely, both occupation states of a topological junction generally carry Josephson currents, with their currents being equal in magnitude but opposite in sign. Thus, both steps should depend on flux in a symmetric manner. This difference is clearly seen in Fig.\ \ref{fig.Psw2}.

\subsection{Poisoning dynamics}
\label{sec:pois}

According to Eq.\ (\ref{eq:Psw_aver}), the measured switching probability is sensitive to the probabilities $c_n$ for the various occupation states $n$ of the junction. As shown experimentally in Ref.\ \cite{zgirski11}, this can be used to extract the poisoning dynamics of the weak Josephson junction by a ``pump-probe" technique. This technique can be readily extended to topological Josephson junctions. 

The basic idea of the technique \cite{zgirski11} is to drive the occupation probabilities $c_n$  out of equilibrium, e.g., by a short initial current pulse, and to probe the switching current by a second current pulse at a later time $t$. With increasing time delay $\Delta t$ between the current pulses, the junction occupations relax back towards equilibrium, and this is reflected in the switching probability $P_{\rm sw}$, due to its dependence on the $c_n$. 

\begin{figure}[t]
 \includegraphics[width=0.45\textwidth]{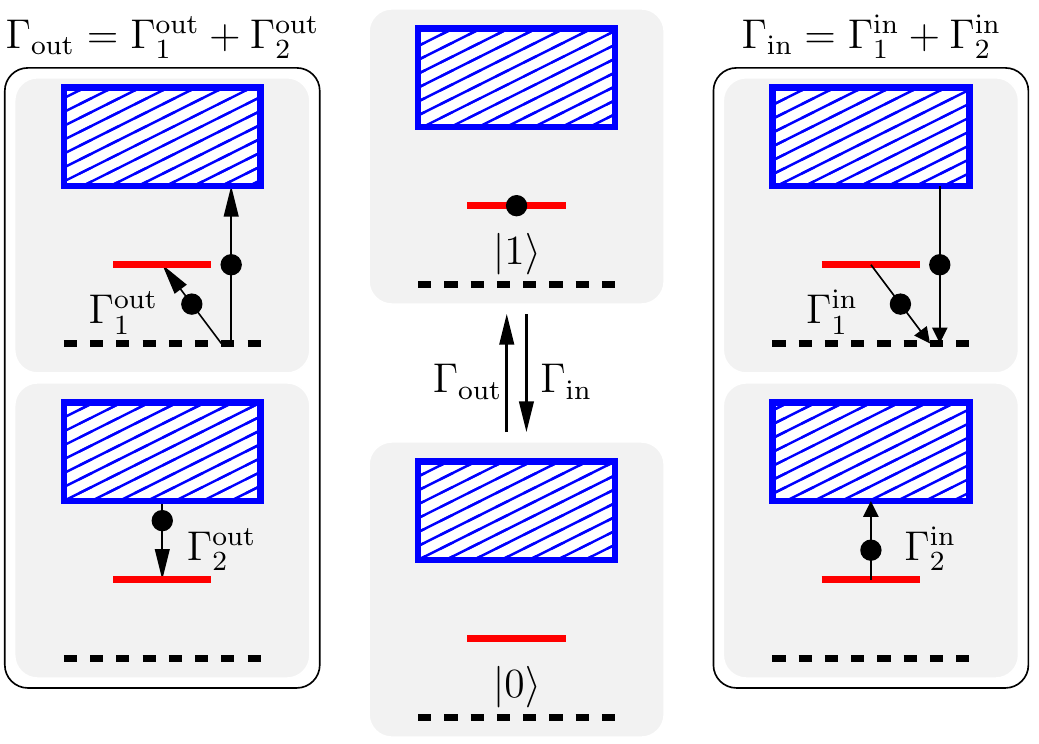}
 \caption{\label{fig.poisoning} Center: Parity switching between states $\ket{1}$ and $\ket{0}$, with rate $\Gamma_{\rm in}$ and $\Gamma_{\rm out}$. Left: Quasiparticle
 processes that contribute to $\Gamma_{\rm out}$. The top panel shows the breaking of a Cooper pair, with one electron excited into the subgap state (red line) and the second electron excited to the continuum (blue box) . The bottom panel shows the transition of a quasiparticle from the continuum into the subgap state. Right: Quasiparticle processes that contribute to $\Gamma_{\rm in}$. The top panel shows the recombination of quasiparticle excitations from the continuum and the subgap state into a Cooper pair. The bottom panel shows the excitation of an excitation from the subgap state into the quasiparticle continuum.}
\end{figure}

This can be used to extract the dependence of the $c_n$'s on the time delay $\Delta t$ and hence the poisoning rates by comparison with a simple rate equation. The dominant poisoning processes in a short topological junction are shown in Fig.\ \ref{fig.poisoning}. Note that in short junctions, the presence of above-gap quasiparticles leaves the Josephson current unchanged. Denoting the occupations of the state $|0\rangle$ and $|1\rangle$ by $p$ and $1-p$, respectively, the rate equation takes the form
\begin{equation}
   \frac{{\mathrm d}p}{{\mathrm d} t} = -\Gamma_{\rm out} p + \Gamma_{\rm in} (1 - p).
\end{equation}
In equilibrium, this is solved by $p=p_\infty= \Gamma_{\rm in}/(\Gamma_{\rm in}+\Gamma_{\rm out})$, and this equilibrium is approached with rate $\Gamma= \Gamma_{\rm in}+\Gamma_{\rm out}$. Both $\Gamma$ and $p_\infty$ can be measured, yielding the poisoning rates $\Gamma_{\rm in}$ and $\Gamma_{\rm out}$. 

While quasiparticle poisoning frequently suppresses Majorana signatures such as the $4\pi$-periodic Josephson effect or the $2e^2/h$ conductance quantization of a Majorana tunnel junction, measurements of the poisoning dynamics may actually be helpful in distinguishing between topological and nontopological junctions. This is related to the fact that a nontopological junction typically has additional channels of poisoning dynamics which are absent in a short topological junction. Specifically, a nontopological junction can have two pairs of subgap states while a topological junction has only one. As a result, we can have poisoning processes in a nontopological junction in which a Cooper pair is split up between (or recombined from) the two positive-energy subgap states. 
No such process exists in a short topological junction where all poisoning processes necessarily involve the quasiparticle continuum, as shown in Fig.\ \ref{fig.poisoning}. 

This difference becomes particularly dramatic and helpful at $\delta =\pi$ when the nontopological junction has only weakly anticrossing Andreev levels. Such a situation is shown in Fig.\ \ref{fig:ABS_change_mu} in Sec.\ \ref{Sec:TJJ_spin}. Then, it may be challenging to resolve the weak splitting in switching-current measurements. However, the poisoning dynamics of the two settings remains distinctly different. The fastest rate for the topological junction has an activated temperature dependence with an activation energy of the order of the topological superconducting gap. In contrast, the fastest rate of a nontopological junction should involve a considerably smaller activation energy which equals the sum of the energies of the spin-up and spin-down Andreev levels. 

\section{Microwave absorption\label{Sec:Theory}}

In addition to the switching current, topological and nontopological Josephson junctions also differ in their microwave absorption. Microwave absorption was studied for
nontopological junctions by Kos {\em et al.} \cite{kos13} and for long topological junctions by V\"ayrynen {\em et al.} \cite{vayrynen2015}. Here, we present corresponding results
for short topological Josephson junctions. (Related results were also obtained very recently in Ref.\ \cite{zazunov2016}.) For definiteness, we consider a model Hamiltonian of a
short topological Josephson junction which is appropriate for a topological Josephson junction based on a proximity-coupled topological-insulator edge \cite{fu09}. This model
allows us to explicitly compute the Josephson current and the transition rates for the various microwave-induced quasiparticle processes.  Related calculations of
    admittance of a topological wires have been done in Ref.~\cite{Dmytruk15,Dmytruk16}.

\subsection{Bound states and Josephson current}
\label{sec:bound}

Consider the Fu-Kane model of a topological Josephson junction \cite{fu09}. The banks, consisting of a topological insulator edge proximity coupled to a conventional superconductor, are separated by a section in which the edge state is gapped out by a Zeeman field or proximity coupling to a ferromagnetic insulator. The banks are considered to be long enough that the Majorana bound states at the junction are decoupled from other Majoranas far from the junction. We also require the edge state to be well separated so that we can focus on an individual edge mode. 

To model a short junction for which the length $L$ of the junction is small compared to the superconducting coherence length, we take the limit $L\to 0$ while keeping $R=ML/v_{F}$ fixed, where $M$ is the strength of the magnetic gap in the junction, i.e., we treat the Zeeman field as a $\delta$-function perturbation. In the Nambu basis $\Psi=(\psi_{\uparrow},\psi_{\downarrow},\psi_{\downarrow}^{\dagger},-\psi_{\uparrow}^{\dagger})^{T}$, the Hamiltonian takes the form $H=\frac{1}{2}\Psi^\dagger \mathcal{H} \Psi$ with
\begin{equation}
    \mathcal{H}=v_{F}p\sigma_{z}\tau_{z}+\Delta(x)\tau_{x}+M(x)\sigma_{x}  
\end{equation}
where $x$ ($p$) denotes the coordinate (momentum) along the topological-insulator edge, $v_F$ is the edge-mode velocity, and $\sigma_j$ and $\tau_j$ are Pauli matrices in spin and Nambu (particle-hole) space, respectively. The proximity-induced superconducting gap 
\begin{eqnarray}
    \Delta(x)&=&\Delta\left[ \theta(-x-L/2)+e^{i\phi\tau_z}\theta(x-L/2)\right] \nonumber\\
   &\to& \Delta e^{i\phi(x)\tau_z} 
\end{eqnarray}
has strength $\Delta$ and a phase difference of $\phi$ across the junction located at $x=0$, so that $\phi(x)=\phi\theta(x)$. (In this section, we use $\phi$ instead of $\delta$ to avoid confusion with the $\delta$-function). Similarly, the magnetic gap takes the form
\begin{equation}
    M(x)= M\theta(x+L/2)\theta(-x+L/2) \to v_F R \delta(x)
\end{equation}
in the short-junction limit. 

Thus, we can also write the Hamiltonian as 
\begin{equation}
    \mathcal{H}= v_{F}p\sigma_z\tau_z +\Delta e^{i\phi(x)\tau_z}\tau_{x}+v_F R\delta(x)\sigma_{x}.
\end{equation}
The spatial dependence of the superconducting phase can be eliminated by a local gauge transformation, $U=e^{i\phi(x)\tau_z/2}$. This transforms the Hamiltonian into
\begin{equation}
    U^\dagger\mathcal{H}U=v_{F} {p} \sigma_{z}\tau_{z}+\Delta\tau_{x}+v_{F}\left[\frac{\phi}{2}\sigma_{z}+R\sigma_{x}\right]\delta(x),
    \label{eq.transformed_H}
\end{equation}
which we will denote as $\mathcal{H}$ in the following. 

The connection formula across the $\delta$-function can be readily derived by rearranging the Bogoliubov-de Gennes equation ${\cal H}\psi = E\psi$ as
\begin{equation}
   i\frac{\partial\psi}{\partial x} = -\frac{1}{v_F}\sigma_z\tau_z\left[ E-\Delta\tau_x -v_F \left(\frac{\phi}{2}\sigma _z +R\sigma_x\right)\delta(x) \right] \psi.
\end{equation}
By writing the solution as $\psi(x) = U(x,x_0) \psi(x_0)$ in terms of the state at some reference point $x_0$, we find
\begin{eqnarray}
  U(x,x_0) &=& {\cal P} \exp\left\{\frac{i}{v_F} \sigma_z\tau_z \int_{x_0}^x {\mathrm d}x^\prime \Big[ E-\Delta\tau_x 
  \right. \nonumber\\
  && \,\,\,\,\, \left. -v_F\left( \frac{\phi}{2} \sigma_z+R\sigma_x\right)\delta(x^\prime)\Big]\right\},
\end{eqnarray}
where ${\cal P}$ is an ordering operator which moves larger $x$ to the left. Specifically, we can now compute 
\begin{equation}
   U(0^+,0^-) = e^{-i\phi\tau_z/2} [\cosh R + \sigma_y\tau_z \sinh R],
\label{eq.connection_condition}
\end{equation}
which connects the states on the two sides of the $\delta$-function, $\psi(0^+)= U(0^+,0^-)\psi(0^-)$.

We can use this connection formula to obtain the bound states localized at the junction. To do so, we match the properly decaying solutions of the Bogoliubov-de Gennes equation on the left and right sides of the $\delta$-function by means of the connection formula (\ref{eq.connection_condition}) and obtain one pair of localized Andreev bound states $\pm E_M(\phi)$ with
\begin{equation}
    E_M(\phi)=\frac{\Delta}{\cosh R }\cos\frac{\phi}{2} = \sqrt{D}\Delta\cos\frac{\phi}{2}.
\label{abs_tj}
\end{equation}
Here, we have defined the junction transmission $D=1/\cosh^{2} R$. This pair of Andreev bound states emerges from the pair of coupled Majorana bound states adjacent to the topological Josephson junction. For completeness, we include details of this calculation in App.\ \ref{app.ABS}. 

Combining Eqs.\ (\ref{eq:In_topological}) and (\ref{abs_tj}), we can obtain the Josephson current as 
\begin{equation}
    \mathcal{I}_n = \frac{e\Delta}{2\cosh R}\sin\frac{\phi}{2}(1-2n)
     = \frac{\pi G}{2}\frac{\Delta^2\sin\phi}{2eE_M(\phi)}(1-2n), \label{eq.current1}
\end{equation}
where $n=0,1$ denotes the occupancy of the bound state and we defined $G=e^2D/\pi$. For a given junction occupation $n$, the Josephson current is $4\pi$-periodic in $\phi$ and the two states of the junction carry exactly opposite supercurrents, as shown in Fig.~\ref{fig.comparison}(b$^\prime$).

\subsection{Linear response to microwave radiation}
\label{sec:admittance}

We model the microwave radiation as an applied time-dependent bias $V(t)$ which modifies the phase difference across the junction according to $\phi\rightarrow\phi-2\phi_{1}(t)$, where $\dot{\phi}_{1}(t)=eV(t)$. We assume that the microwave radiation of frequency $\omega$ is weak, $\phi_{1}\sim \abs{eV/\omega}\ll 1$, so that we can treat the perturbation 
\begin{align}
    H'(t')&=v_{F}\left[\psi_{+}^{\dagger}(0)\psi_{+}(0)-\psi_{-}^{\dagger}(0)\psi_{-}(0)\right]\phi_{1}(t')\nonumber\\
    &=\frac{1}{e}I(t')\phi_{1}(t'),
\end{align}
in linear response. We note in passing that we neglect the shift in chemical potential by $eV(t)$. This term yields a purely real response function and is thus irrelevant for microwave absorption \cite{kos13}.
 
Using the Kubo formula, the current response to the microwave radiation can be expressed as
\begin{align}
    \delta\braket{I(t)}&=-i\int_{-\infty}^{t}\braket{[I(t),H'(t')]}dt'\nonumber \\
    &=-\frac{i}{e}\int_{-\infty}^{t}\braket{[I(t),I(t')]}\phi_{1}(t')dt',
\end{align}
and described by the response function
\begin{equation}
    \chi(t)=-\frac{i}{e}\theta(t)\left\langle [I(t),I(0)]\right\rangle.
    \label{eq.chi}
\end{equation}
The admittance $Y(\omega)$ of the junction can be written as $Y(\omega)=\frac{ie}{\omega}\chi (\omega)$, where $\chi(\omega)$ denotes the Fourier transform of $\chi(t)$. The linear absorption rate $W$ of the microwave radiation becomes \cite{kos13} 
\begin{equation}
    W=\frac{\phi_1^2}{2e^2}\omega {\rm Re}Y(\omega), \quad \omega>0.
\label{absorptionrate}
\end{equation}
This quantity is a measure of the microwave-induced rate of change of the weight factors $c_n$ in $P_{\rm sw}$ as given in Eq.\ (\ref{eq:Psw_aver}).

The admittance can be computed by using the current operator
\begin{equation}
    {I}=ev_{F}\left[\psi_{+}^{\dagger}(0)\psi_{+}(0)-\psi_{-}^{\dagger}(0)\psi_{-}(0)\right], \label{eq.current}
\end{equation}
where $\psi_{\pm}(0)$ is the annihilation operator for the left/right moving electron at position $x=0$ of the junction. We need to choose either $x=0^+$ or $0^-$ for the wavefunctions to be well defined.
The electron operators can be expressed in terms of the Bogoliubov quasiparticle operators $\gamma_{\nu}$ \cite{badiane11},
\begin{gather}
      \psi_{+}(0)=\sum_{\nu}u_{+\nu}(0)\gamma_{\nu}-v_{-\nu}^{*}(0)\gamma_{\nu}^{\dagger}  \nonumber \\
      \psi_{-}(0)=\sum_{\nu}u_{-\nu}(0)\gamma_{\nu}+v_{+\nu}^{*}(0)\gamma_{\nu}^{\dagger}. \label{eq.bogoliubov}
\end{gather}
Here, we introduced the spinor wave functions $\Psi_\nu=(u_{+\nu},u_{-\nu},v_{+\nu},v_{-\nu})$. The Andreev bound state is labeled by $\nu=0$ and the continuum states by $\nu=(E,\eta,\chi)$, with $\eta=e,h$ and $\chi=l,r$ corresponding to the state generated by incoming electron/hole states from the left/right. The $\pm$ label refers to the two spin components which are locked to the propagation directions of the edge channel. 

\begin{figure}[t]
\includegraphics[width=0.45\textwidth]{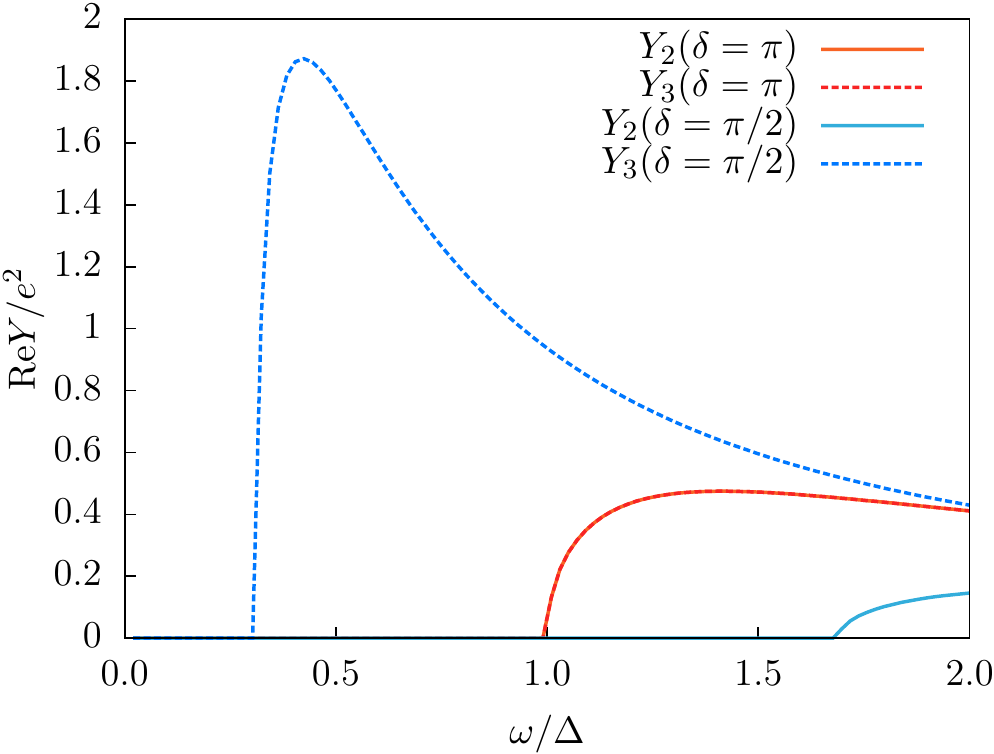}
\caption{\label{fig.admittance} Various contributions to the real part of the admittance for the Fu-Kane model, based on Eq.\ (\ref{eq.ReAd}), for $D=0.95$  and phase differences $\phi=\pi$ as well as $\phi=\pi/2$. For $\phi=\pi$, $E_A=0$, so that ${\rm Re}Y_2$ and ${\rm Re}Y_3$ coincide. At phase differences $\phi$ away from $\pi$, the two curves differ.}
\end{figure}

By using the explicit expressions for the wave functions of both bound and continuum states, as calculated in App.\ \ref{app.wf}, we can first recover the Josephson current given in Eq.\ (\ref{eq.current1}). The corresponding  derivation is given in App.\ \ref{app.JC}. Extending the calculation to the current-current correlation function (\ref{eq.chi}), we can then obtain microscopic results for the admittance of short Josephson junctions, as shown in App.\ \ref{app.resp}. We neglect above-gap excitations, as they are suppressed by the superconducting gap. Then, the real part of the admittance can be written as a sum of three terms,
\begin{equation}
{\rm Re}Y={\rm Re}Y_1 + (1-n){\rm Re}Y_2+ n{\rm Re}Y_3.
\label{eq.Y2Y3}
\end{equation}
The three terms correspond to three different quasiparticle processes shown in Fig.\ \ref{fig.process}(a$^\prime$). Explicitly, ${\rm Re}Y_1(\omega)\propto \theta(\omega-2\Delta)$ corresponds to the process (1) in which a Cooper pair is excited into the continuum as two quasiparticles. This process requires a threshold energy of $2\Delta$. ${\rm Re}Y_2(\omega)\propto\theta(\omega-\Delta-E_M)$ describes the process (2), in which a Cooper pair is split between the Andreev level and the quasiparticle continuum. This process requires a threshold energy $\Delta+E_M$ and an initially empty Andreev level. Finally,  ${\rm Re}Y_3(\omega)\propto \theta(\omega-\Delta+E_M)$ corresponds to the process (3), in which a quasiparticle is excited from the Andreev level into the continuum. This requires a threshold energy of $\Delta-E_M$ and an occupied Andreev level. Unlike for conventional Josephson junctions as discussed in Ref.\ \cite{kos13}, there is no process with absorption energy $2 E_M$ as the Andreev level is nondegenerate for a topological Josephson junction. 

\begin{figure}[t]
    \includegraphics[width=0.45\textwidth]{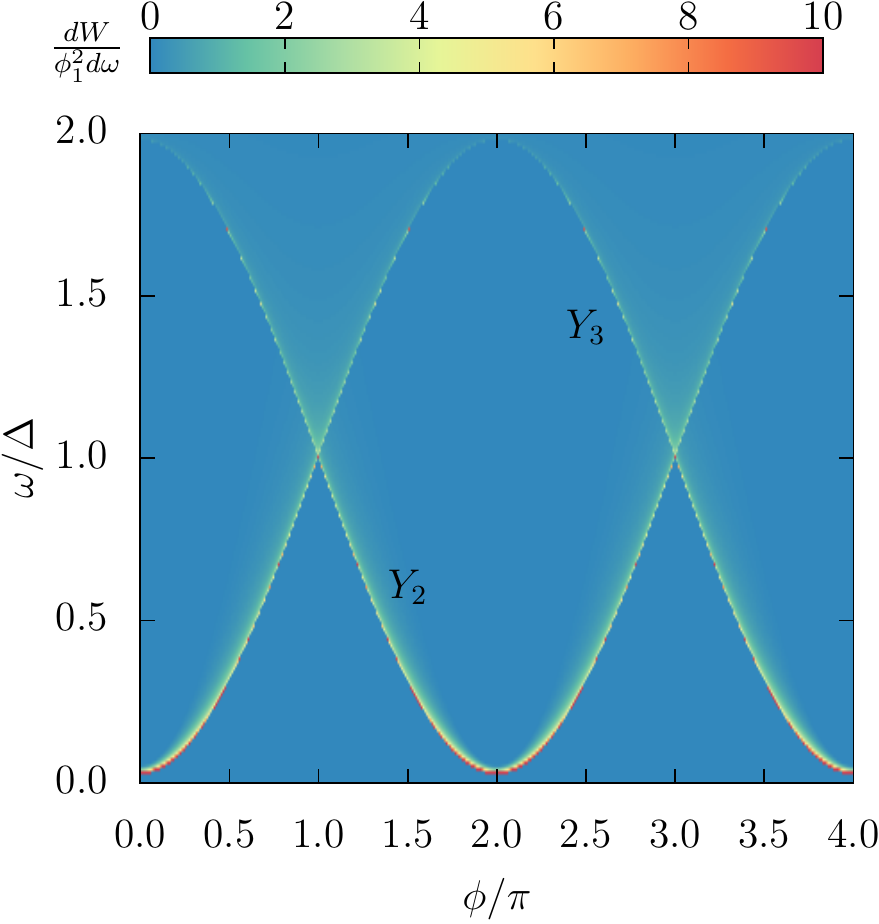}
    \caption{\label{fig.dY-dw} Derivative of the linear absorption rate with respect to the microwave frequency, ${\mathrm d}W/{\mathrm d}\omega$, (see Eq.\ \ref{absorptionrate}). For optimal visibility of the thresholds, we assume an occupation of $n=1/2$ in Eq.~(\ref{eq.Y2Y3}) independently of flux. While the figure displays the sum of contributions from $Y_2$ and $Y_3$, the bright curves result predominantly from $Y_2$ and $Y_3$ as labeled in the figure. }
\end{figure}

Detailed expressions for these functions are included in Eq.\ (\ref{eq.ReAd}) in App.\ \ref{app.resp}. The explicit expressions show that the thresholds at $\Delta+E_M$ and $\Delta-E_M$ are sharp in the sense that their derivatives with respect to $\omega$ have square-root singularities at the threshold. This is shown in Fig.\ \ref{fig.admittance}, which plots ${\rm Re}Y_2$ and ${\rm Re}Y_3$ for various phase differences $\phi$ across the junction. These results also allow one to compute the absoption rate ${\mathrm d}W/{\mathrm d}\omega$ according to Eq.\ (\ref{absorptionrate}). A corresponding color plot as a function of both $\phi$ and $\omega$ which emphasizes the threshold energies is shown in Fig.\ \ref{fig.dY-dw}. Here, we assume that both parity states are equally populated, independently of the applied flux. 

\section{Topological {\em vs} nontopological junctions: Effects of Zeeman field and spin-orbit coupling \label{Sec:TJJ_spin}}

Potential realizations of topological Josephson junctions require systems which involve spin-orbit coupling and/or Zeeman fields. When searching for topological superconductivity, one is thus dealing with Josephson junctions which are subject to both of these. Strictly speaking, our considerations for nontopological junctions in the previous sections did not include these effects. One may thus worry that their inclusion makes the proposed experimental distinctions between topological and nontopological junctions less clear-cut. This question is addressed in the present section. Our principal conclusion is that the signatures remain essentially robust as long as one considers short Josephson junctions. 

Important realizations of topological Josephson junctions rely on 2d topological insulators \cite{fu09} or semiconductor quantum wires \cite{lutchyn10,oreg10}. In Sec.\ \ref{sec:bound}, we presented microscopic results for short junctions made of 2d topological insulator edges, subject to a Zeeman field in the junction region. These junctions are topological, and their subgap spectrum agrees with the generic subgap spectrum of short topological junctions which underlies the considerations of this paper. At the same time, there is experimental evidence that there can be edge-state transport even in the trivial regime \cite{nichele2015}. For this reason, in Sec.\ \ref{sec:B_inside}, we study short nontopological junctions which are one-dimensional and subject to a strong Zeeman field inside the junction region. In Sec.\ \ref{rashbaJJ}, we explore Josephson junctions based on proximity-coupled semiconductor quantum wires with Zeeman and spin-orbit coupling.  This model can be explicitly tuned between the topological and the nontopological phase.

\subsection{Nontopological Josephson junctions with strong Zeeman field in the junction region}
\label{sec:B_inside}

\begin{figure}[t]
  \includegraphics[width=0.45\textwidth]{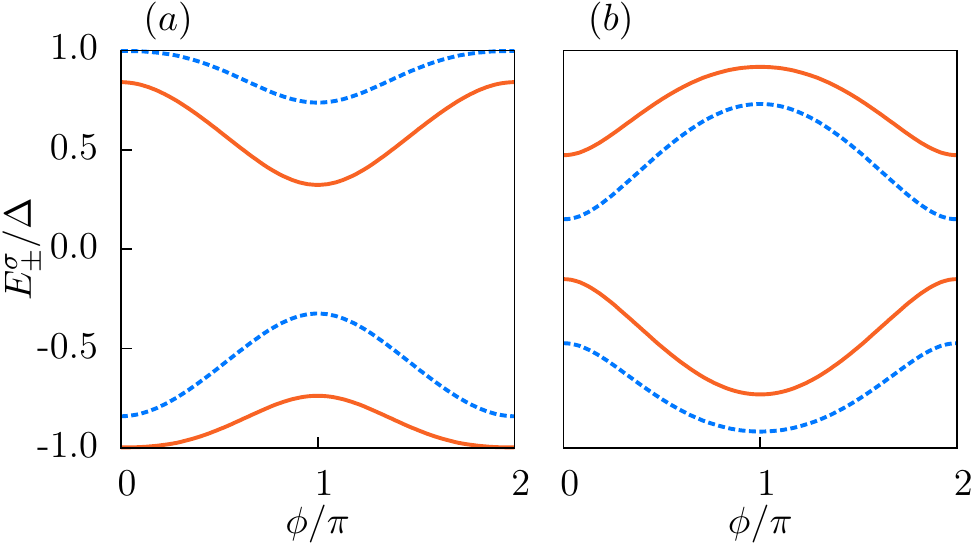}
  \caption{\label{fig.ABS_spin_BdG} Subgap energies of a short conventional Josephson junction as a function of the phase difference, in the presence of Zeeman field applied in the junction region. The orange solid curves are the spectra for spin up with Nambu spinor $(\psi_{\uparrow},\psi_{\downarrow}^\dagger)^T$. The blue dashed lines are the the corresponding spectra for spin down follow from particle-hole symmetry. The panels illustrate the two types of typical behaviors, with parameters chosen as (a) $\tilde{\eta}=0.5$ and (b) $\tilde{\eta}=2.8$, with $D=0.6$ and $R\cos\tilde{\gamma}=0.2$ in both panels.}
\end{figure}

Consider a Josephson junction made from a nontopological (i.e., non-helical) one-dimensional channel. In the short junction limit, the splitting of Andreev levels due to spin orbit coupling is of order $\Delta^{2} \tau_{\rm dw}/\hbar$\cite{Michelsen08,Chtchelkatchev03hjk,Beri2008}. Here, $\tau_{\rm dw}$ denotes the dwell time in the junction which approaches zero in the short-junction regime. Hence, we can neglect spin-orbit coupling and focus on the Zeeman field. The subgap states and Josephson current of such junction can quite generally be obtained by scattering theory \cite{beenakker91}, see App.~\ref{app.B_inside} for a detailed calculation. Figure \ref{fig.ABS_spin_BdG} shows two typical subgap spectra as a function of the phase difference across the junction in the case of a short nontopological junction with Zeeman field inside the junction region. 

We observe that in short junctions, the main consequence of Zeeman and spin-orbit coupling in the junction region is that the odd-parity state spin splits. This implies that the odd-parity states carry nonzero supercurrent, leading to an additional plateau in the switching probability. This actually enhances the contrast with the short topological junctions which exhibit a single plateau. Even if this additonal plateau is not resolved, however, we find that the supercurrent still vanishes when $\phi$ is an odd multiple of $\pi$. Thus, the behavior of the plateau width with phase difference remains as discussed in Sec.\ \ref{Sec:Psw}. 

The spin splitting of the odd-parity states also modifies the behavior in microwave absorption. Let us denote the two positive-energy Andreev levels as $E_\pm$. Then, transitions appear when the microwave frequency equals (i) $E_{+}+E_{-}$, generalizing the line at $2E_A$ in the absence of the Zeeman field, (ii) $\Delta\pm E_{+}$ or $\Delta \pm E_{-}$, generalizing the  lines at $\Delta \pm E_A$ to the spin-split case, and (iii) $E_{+}-E_{-}$. The latter is visible only due to spin-orbit coupling and should therefore be weak. Thus, the magnetic field and spin-orbit coupling introduce additional absorption lines in microwave absorption, while short topological Josephson junctions have only two aborption lines. 

\subsection{Josephson junctions based on proximitized Rashba nanowires}
\label{rashbaJJ}

Nontopological junctions based on proximity-coupled Rashba nanowires include both Zeeman and spin-orbit coupling also in the superconducting leads. Here, we explore the corresponding modifications for short junctions and show that both switching-current and $ac$-absorption measurements continue to provide clear-cut distinctions between topological and nontopological junctions.

\begin{figure}[t]
    \includegraphics[width=0.48\textwidth]{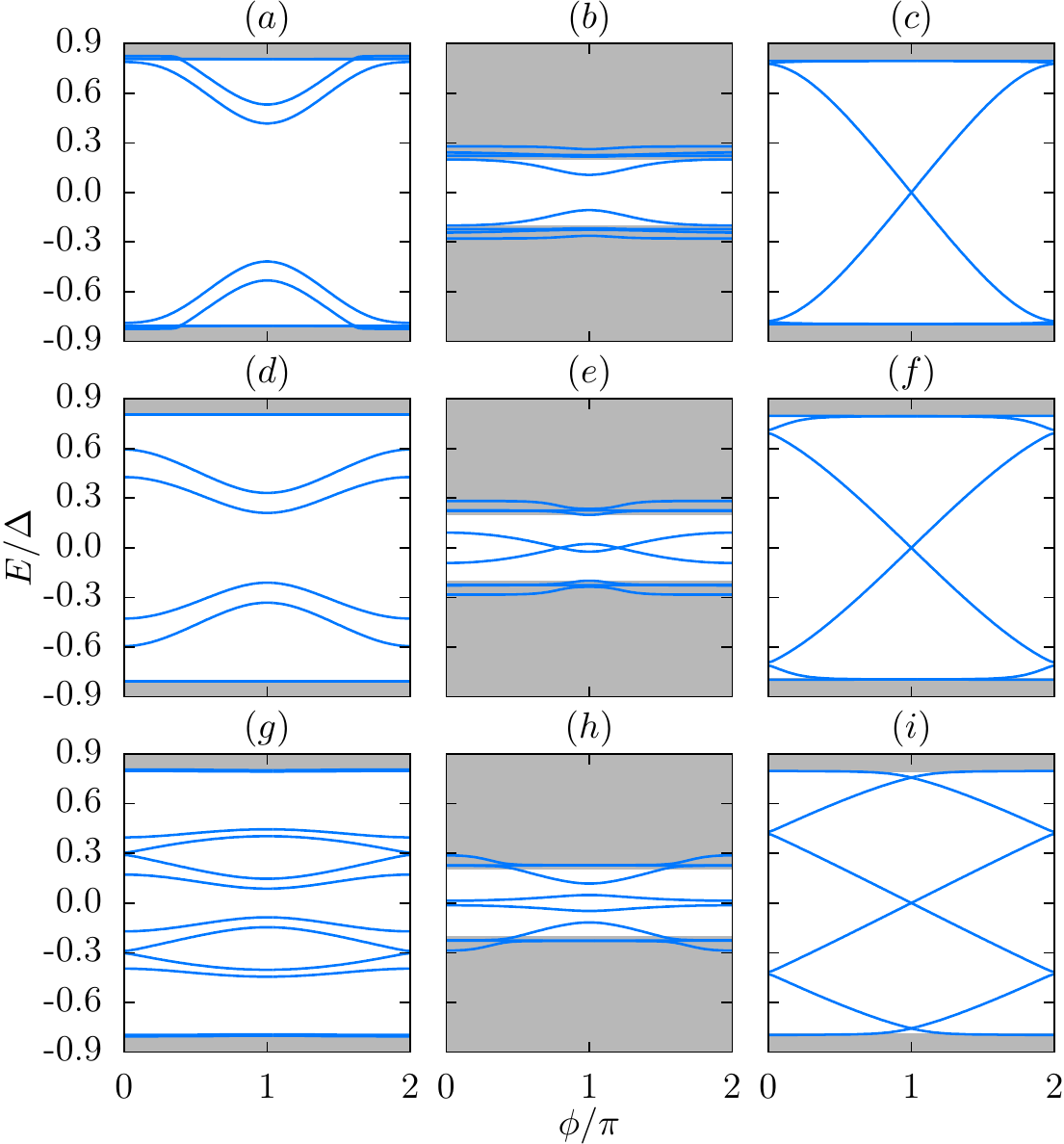}
    \caption{\label{fig:ABS_wire} Low-energy spectra of Hamiltonian (\ref{eq:wire}) as a function of phase difference $\varphi$, for various Zeeman fields and junction lengths. The results are obtained numerically for finite-length samples, showing all levels which become subgap states at least for some range of phase differences. Energies corresponding to the quasiparticle continuum of infinite wires are shown in grey. We choose a chemical potential $\mu = 0$, spin-orbit interaction $m\alpha^2 = \Delta$ and a total length
    $60\xi$ of the system, with $\xi=2\alpha/\Delta$ the bulk coherence length of the superconductor when $B=0$. Results for a short junction with $L = 0.05\xi$ are shown in (a)-(c) for increasing Zeeman field: (a) nontopological junction, $B=0.2\Delta$; (b) nontopological junction, $B=0.8\Delta$; (c) topological junction, $B=2.0\Delta$. The subgap spectrum behaves in a qualitatively similar manner in intermediate length junctions with $L =0.5\xi$. Results for junctions of this length are shown in Figs.\ (d)-(f), with the other parameters equal to those of panels (a)-(c). Additional subgap states emerge only in long junctions, as shown in panels (g)-(h) for $L =2\xi$, and other parameters again as in (a)-(c). The numerical results are obtained by discretizing the Hamiltonian (\ref{eq:wire}) with a minimal spacing of $0.025\xi$ and an eighth-order approximation to the Laplacian.  }
\end{figure}

\begin{figure}[t]
    \includegraphics[width=0.48\textwidth]{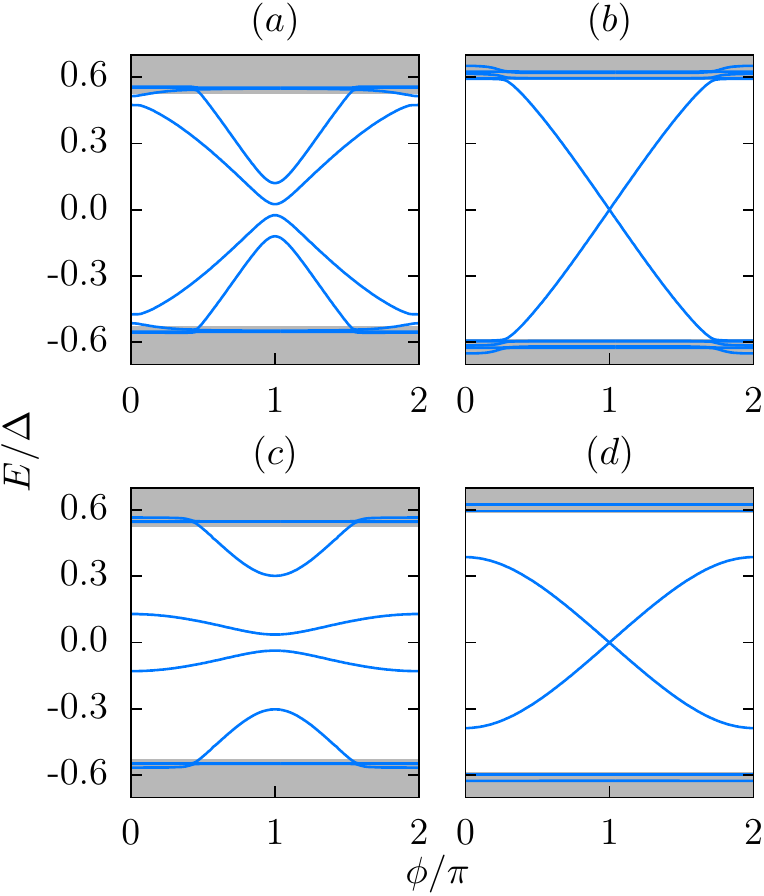}
    \caption{\label{fig:ABS_change_mu} Low-energy spectra of Hamiltonian (\ref{eq:wire}) as a function of phase difference $\varphi$, for fixed  Zeeman fields $B=2.0\Delta$ and junction length $L=0.5\xi$. The results are obtained numerically for finite-length samples, showing all levels which become subgap states at least for some range of phase differences. Energies corresponding to the quasiparticle continuum of infinite wires are shown in grey. We choose a spin-orbit interaction of $m\alpha^2 = \Delta$ and a total length of $60\xi$ of the system, with $\xi=2\alpha/\Delta$ the bulk coherence length of the superconductor when $B=0$. (a) Nontopological junction with $\mu = 3.0\Delta$. (b) Topological junction with $\mu = 1.0\Delta$. Panels (c) and (d) are for parameters as in (a) and (b), respectively, but with an additional potential barrier of height $3\Delta$ in the junction region, which reduces the junction transmission. The numerical results are obtained by discretizing the Hamiltonian (\ref{eq:wire}) with a minimal spacing of $0.025\xi$ and an eighth-order approximation to the Laplacian.}
\end{figure}

The explict Hamiltonian and the bulk dispersions for this system are given by Eq.~(\ref{eq:wire}) in App.~\ref{app.wire}. We compute the spectrum of the Hamiltonian (\ref{eq:wire}) numerically by discretizing the model into a finite difference representation. The results are shown in Figs.~\ref{fig:ABS_wire} and \ref{fig:ABS_change_mu}. In Fig.~\ref{fig:ABS_wire}, we fix the chemical potential to $\mu=0$. Results for a short junction are shown in panels (a)-(c), with the Zeeman field increasing from (a) to (c). Far on the nontopological side of the topological phase transition [panel (a)], the results differ from those for the simplified model of a nontopological junction in that the subgap states are spin split, leading to four subgap states. As argued in the previous section \ref{sec:B_inside}, this leads to additional plateaus in the switching probability and additional lines in microwave absorption, which enhances the central distinctions between short topological and nontopological junctions. 

When approaching the topological phase transition by increasing the Zeeman field, the bulk gap $\Delta - B$ becomes smaller and two of the subgap states merge with the continuum. This is shown in panel (b) of Fig.~\ref{fig:ABS_wire}. Thus, the switching probability is expected to exhibit only a single plateau, as in the topological phase. However, the plateau width remains distinctly different as the Josephson current vanishes at $\phi = \pi$ where it becomes maximal in a topological junction. The latter can be seen from panel (c) which shows the subgap spectrum in the topological phase. 

The results in panels (a)-(c) of Fig.\ \ref{fig:ABS_wire} were obtained for a very short junction with a length of $L=0.05\xi$, where $\xi$ is the superconducting coherence length for $B=0$ and $\mu=0$. Qualitatively the same results are found for intermediate length junctions with $L=0.5\xi$, as shown in panels (d)-(f). Additional subgap states appear only for even longer junctions of length $L=2\xi$, as shown in panels (g)-(i) of Fig.\  \ref{fig:ABS_wire}. 

We can also tune the junction across the topological phase transition by varying the chemical potentials $\mu$. Corresponding results are shown in Fig.~\ref{fig:ABS_change_mu} for a junction of moderate length, $L=0.5\xi$. Panel (a) corresponds to a nontopological junction with spin-split subgap states and vanishing supercurrent for $\phi=\pi$. Panel (b) corresponds to a topological junction with maximal supercurrent at $\phi=\pi$. Potential scattering in the junction region reduces the junction transmission which opens a gap between subgap states and quasiparticle continuum in the topological case while leaving the behavior near $\phi=\pi$ qualitatively unchanged. Corresponding numerical subgap spectra are shown in Fig.\ \ref{fig:ABS_change_mu}, see panels (c) and (d).

\section{Conclusion\label{Sec:con}}

The most immediate detection of a topological Josephson junction relies on the $4\pi$-periodic Josephson effect. Its observation is, however, complicated by quasiparticle poisoning and diabatic effects. In this paper, we showed that topological Josepshon junctions can be identified even in the presence of quasiparticle poisoning. The proposed techique relies on switching current measurements. While such measurements may be interesting even for the Josephson junction by itself, they provide much more information when including the junction into an asymmetric SQUID, together with an auxiliary junction with much larger critical current. Most importantly, incorporation into an asymmetric SQUID allows for phase-resolved measurements of the Josephson junction of interest. 

Rather than measuring the $4\pi$ periodicity of individual subgap levels, the proposed switching current measurements probe the existence of a protected level crossing at a certain phase difference $\delta$. While in a topological junction, this level crossing is protected by fermion parity, there is no corresponding protection in nontopological junctions. For a particular junction state, the Josephson current is correspondingly maximal in magnitude at the protected crossing of a topological junction, but vanishes in a nontopological junction. We showed that this has striking manifestations in the switching probability of the junction as a function of the height of the applied current pulse, as illustrated in Figs.\ \ref{fig.Psw1} and \ref{fig.Psw2}. 

Specifically, our considerations focused on short Josephson junctions for which the number of subgap states is limited and the differences between topological and nontopological junctions are most pronounced. Especially, near degeneracies of subgap levels are quite unlikely in short nontopological junctions, as we show by explicitly calculating the subgap spectra for specific models based on topological-insulator edge modes or semiconductor quantum wires. 

The prosposed measurements are not only tolerant of (and in fact exploit) quasiparticle poisoning, but also provide access to the poisoning dynamics. The poisoning rates can be extracted by means of a pump-probe technique with multiple current pulses offset in time. As we showed, this is particularly useful to identify nontopological junctions with anomalously weak anticrossings of the Andreev levels. Microwave irradiation may be another useful technique in probing the poisoning dynamics as it also drives the system out of equilibrium. Moreover, microwave absoption provides access to the subgap spectrum of Josephson junctions, providing additional signatures which differentiate topological from nontopological junctions.

Thoughout our discussion, we focused on the Majoranas which are localized at the junction and did not consider additional Majoranas located far from the junction. This is justified when the overlap between these additional outer Majoranas and the junction Majoranas can be neglected. Then, the subgap states resulting from the outer Majoranas are independent of the phase difference across the junction and the Josephson currents remains unaffected. Similarly, the transition matrix elements vanish for microwave processes involving both these and the junction Majoranas. 

It is interesting to consider how our results become modified when there is substantial overlap between the junction and outer Majoranas. A topological junction would now have a ``counterpart'' of process (4) in Fig.\ \ref{fig.process},  yielding an absorption line at the energy $E_M$ plus the small splitting of the outer Majoranas (as opposed to $2E_A$ for a nontopological junction). In addition, there should also be an absorption feature with a threshold near $\Delta$, which distinguishes this case from conventional Andreev states. Both additional features should be much dimmer than other features as they require overlap of the outer and junction Majoranas. In switching current measurements, the signature in Fig.\ \ref{fig.Psw1}(b) weakens a bit: In an exponentially narrow window around $\phi=\pi$, the plateau width would go to zero even in the topological phase. The signature in Fig.\ \ref{fig.Psw2}, panels (a) and (b), would only be weakly affected. In particular, the fact that in (b) the plateau in $P_{\rm sw}$ is centered around the same current, should be quite robust.

Thus, we conclude that the proposed signatures remain quite useful in the presence of weak coupling to the outer Majoranas. An exception is the discussion at the end of Sec.\ \ref{sec:pois} concerning poisoning processes. With coupling to outer Majoranas, the activation energy of the poisoning rates of topological junctions would no longer be necessarily larger than the gap, and a measurement without this overlap has clear benefits.

Combined switching current and microwave absorption measurements on the same Josephson junction should thus be a powerful combination to identify topological Josephson junctions. In view of the fact that corresponding measurements have already been carried out successfully on nontopological junctions based on atomic point contacts \cite{Rocca07,zgirski11,bretheau13,bretheau13nature}, we hope that the proposed measurements can be readily implemented for topological junctions. 

\acknowledgments 

We acknowledge discussions with L.\ Glazman and T.\ Karzig, and thank H. Pothier, D. van Woerkom, and R. Egger for comments on the manuscript. This work was supported in part by Priority Programs 1666 and 1285 as well as CRC 183 of the Deutsche Forschungsgemeinschaft (DFG), by the STC Center for Integrated Quantum Materials under NSF Grant No.\ DMR-1231319, by the Israeli Science Foundation (ISF), Minerva grants, a Career Integration Grant (CIG), a Minerva ARCHES prize, and an ERC Grant No. 340210 (FP7/2007-2013).

\appendix
\section{Calculation of wavefunctions\label{app.wf}}
In this appendix, we derive the wave functions for the bound states (App.~\ref{app.ABS}) and continuum states (App.~\ref{App.cont}) of the Fu-Kane model in the short junction limit. The Hamiltonian is given in Eq.~(\ref{eq.transformed_H}).
\subsection{Andreev bound state with $\abs{E} <\Delta$\label{app.ABS}}

For $E<\Delta$, we write down the left ($x<0$) and right ($x>0$) wavefunction which are solutions to the Hamiltonian ${\cal H}$ without the $\delta$-function term,
\begin{gather}
    \Psi_{L}(x)=e^{\kappa x}(aA_0,B_0,A_0,aB_0)^{T}, \quad x<0 \nonumber\\
    \Psi_{R}(x)=e^{-\kappa x}(C_0,aD_0,aC_0,D_0)^{T}, \quad x>0,
\end{gather}
where $\kappa(E)=\sqrt{\Delta^{2}-E^{2}}/v_{F}$ and $a(E)=E/\Delta-i\sqrt{1-E^{2}/\Delta^{2}}$.  
The connection condition in Eq.\ (\ref{eq.connection_condition}) leads to 
\begin{gather}
    e^{i\phi/2}C_0=aA_0\cosh R-i\sinh RB_0 \nonumber \\
    e^{-i\phi/2}C_0=a^{-1}A_0\cosh R+i\sinh RB_0 \nonumber \\
    e^{i\phi/2}D_0=i\sinh RA_0+a^{-1}\cosh RB_0 \nonumber \\
    e^{-i\phi/2}D_0=-i\sinh RA_0+a\cosh RB_0,
\end{gather}
which can be further simplified to become
\begin{gather}
    C_0=\frac{E}{\Delta}A_0\frac{\cosh R}{\cos\phi/2}=A_0 \nonumber \\
     D_0=\frac{E}{\Delta}B_0\frac{\cosh R}{\cos\phi/2}=B_0 \nonumber \\
      C_0=-\frac{\sqrt{D}\sinh RB_0}{\sqrt{1-D\cos^{2}\frac{\phi}{2}}+\sqrt{D}\sin\frac{\phi}{2}}.
\end{gather}
These equations are a set of homogeneous linear equations for the coefficients $A_0$, $B_0$,  $C_0$ and $D_0$. The condition to have nonzero solutions leads to the Andreev bound state energy
\begin{equation}
    E_M(\phi)=\Delta\cos(\frac{\phi}{2})/\cosh  R =\sqrt{D}\Delta\cos(\frac{\phi}{2})
\end{equation}
given in Eq.\ \ref{abs_tj}.   

To obtain the bound state wavefunction, we note that the coefficients fulfill $A_0=C_0$ and
\begin{equation}
    B_0=D_0=\left(\frac{\sqrt{1-D\cos^{2}\frac{\phi}{2}}+\sqrt{D}\sin\frac{\phi}{2}}{\sqrt{1-D\cos^{2}\frac{\phi}{2}}-\sqrt{D}\sin\frac{\phi}{2}}\right)^{1/2}A_0.
\end{equation}
Imposing the normalization condition
\begin{equation}
    \int dx\Psi^{\dagger}(x)\Psi(x)=\frac{2}{\kappa}(\abs{A_0}^{2}+\abs{B_0^{2}})=1,
\end{equation}
and using
\begin{equation}
    \abs{A_0}^{2}+\abs{B_0}^{2}=\frac{2\sqrt{1-D\cos^{2}\frac{\phi}{2}}}{\sqrt{1-D\cos^{2}\frac{\phi}{2}}-\sqrt{D}\sin\frac{\phi}{2}}\abs{A_0}^{2}
\end{equation}
as well as
\begin{equation}
    \kappa(E_{M})=\frac{\Delta}{v_{F}}\sqrt{1-D\cos^{2}\frac{\phi}{2}},
\end{equation}
we obtain 
\begin{gather}
    \abs{A_0}^{2}=\frac{\Delta}{4v_{F}}\left(\sqrt{1-D\cos^{2}\frac{\phi}{2}}-\sqrt{D}\sin\frac{\phi}{2}\right) \nonumber \\
    \abs{B_0}^{2}=\frac{\Delta}{4v_{F}}\left(\sqrt{1-D\cos^{2}\frac{\phi}{2}}+\sqrt{D}\sin\frac{\phi}{2}\right).
\end{gather}

\subsection{Continuum state with $\abs{E}\geq \Delta$\label{App.cont}}

\begin{widetext}
For $E\geq\Delta$, we have four kinds of wave functions $\Psi_{E}^{(\eta,\chi)}$,
 \begin{subequations}
    \label{eq.contwf}
    \begin{align}
        \Psi_{E}^{(e,l)}&=J(E)\left[e^{ipx}(1,0,a,0)+e^{-ipx}(aA^{(e,l)},0,A^{(e,l)},0)^{T}+e^{-ipx}(0,B^{(e,l)},0,aB^{(e,l)})^{T}\right]\theta(-x) \nonumber \\
        &+J(E)\left[e^{ipx}(C^{(e,l)},0,aC^{(e,l)},0)+e^{ipx}(0,aD^{(e,l)},0,D^{(e,l)})\right]\theta(x) 
    \end{align}
    \begin{align}
        \Psi_{E}^{(h,l)}&=J(E)\left[e^{ipx}(0,a,0,1)+e^{-ipx}(aA^{(h,l)},0,A^{(h,l)},0)^{T}+e^{-ipx}(0,B^{(h,l)},0,aB^{(h,l)})^{T}\right]\theta(-x) \nonumber \\
        &+J(E)\left[e^{ipx}(C^{(h,l)},0,aC^{(h,l)},0)+e^{ipx}(0,aD^{(h,l)},0,D^{(h,l)})\right]\theta(x) 
    \end{align}
    \begin{align}
        \Psi_{E}^{(e,r)}&=J(E)\left[e^{-ipx}(aA^{(e,r)},0,A^{(e,r)},0)^{T}+e^{-ipx}(0,B^{(e,r)},0,aB^{(e,r)})^{T}\right]\theta(-x)\nonumber\\
        &+J(E)\left[e^{-ipx}(0,1,0,a)^{T}+e^{ipx}(C^{(e,r)},0,aC^{(e,r)},0)+e^{ipx}(0,aD^{(e,r)},0,D^{(e,r)})\right]\theta(x)
    \end{align}
    \begin{align}
        \Psi_{E}^{(h,r)}&=J(E)\left[e^{-ipx}(aA^{(h,r)},0,A^{(h,r)},0)^{T}+e^{-ipx}(0,B^{(h,r)},0,aB^{(h,r)})^{T}\right]\theta(-x)\nonumber \\
        &+J(E)\left[e^{-ipx}(a,0,1,0)^{T}+e^{ipx}(C^{(h,r)},0,aC^{(h,r)},0)+e^{ipx}(0,aD^{(h,r)},0,D^{(h,r)})\right]\theta(x) 
    \end{align}
\end{subequations}
where $\eta=e,h$ denote electron or hole source, $\chi=l,r$ denote the source field coming from left or right, 
\begin{equation}
    p(E)=\frac{\sqrt{E^{2}-\Delta^{2}}}{v_{F}}, \quad
    a(E)=\frac{E}{\Delta}-\sqrt{\frac{E^{2}}{\Delta^{2}}-1},
\end{equation}
and $J(E)=\left[2\pi v_F(1-\abs{a}^2)\right]^{-1/2}$ is the normalization constant.
These coefficients for the continuum wave functions can be obtained by using the connection condition in Eq.\ (\ref{eq.connection_condition}), 
which will be shown in the following.

\subsubsection{Electron source from left}
For electron source from left, the wavefunction fulfills
    \begin{equation}
        \left(\begin{array}{c}
            C^{(e,l)}\\
           aD^{(e,l)}\\
           aC^{(e,l)}\\
            D^{(e,l)}
        \end{array}\right)=\left(\begin{array}{cccc}
            e^{-i\phi/2}\cosh R & -e^{-i\phi/2}i\sinh R & 0 & 0\\
            e^{-i\phi/2}i\sinh R & e^{-i\phi/2}\cosh R & 0 & 0\\
            0 & 0 & e^{i\phi/2}\cosh R & e^{i\phi/2}i\sinh R\\
            0 & 0 & -e^{i\phi/2}i\sinh R & e^{i\phi/2}\cosh R
        \end{array}\right)\left(\begin{array}{c}
            1+aA^{(e,l)}\\
            B^{(e,l)}\\
            a+A^{(e,l)}\\
            aB^{(e,l)}
        \end{array}\right)
    \end{equation}
\begin{gather}
    A^{(e,l)}=\mathcal{A}=\frac{E(E_{M}^{2}-\Delta^{2})-iE_{M}\sqrt{E^{2}-\Delta^{2}}\Delta\sqrt{D}\sin\frac{\phi}{2}}{\Delta(E^{2}-E_{M}^{2})} \nonumber \\
    B^{(e,l)}=\mathcal{B}=\frac{-iE\sqrt{E^{2}-\Delta^{2}}\tanh R}{E^{2}-E_{M}^{2}}.
\end{gather}
\subsubsection{Hole source from left}
For hole source from left, the wavefunction fulfills
\begin{equation}
    \left(\begin{array}{c}
        C^{(h,l)}\\
        aD^{(h,l)}\\
        aC^{(h,l)}\\
        D^{(h,l)}
    \end{array}\right)=\left(\begin{array}{cccc}
        e^{-i\phi/2}\cosh R & -e^{-i\phi/2}i\sinh R & 0 & 0\\
        e^{-i\phi/2}i\sinh R & e^{-i\phi/2}\cosh R & 0 & 0\\
        0 & 0 & e^{i\phi/2}\cosh R & e^{i\phi/2}i\sinh R\\
        0 & 0 & -e^{i\phi/2}i\sinh R & e^{i\phi/2}\cosh R
    \end{array}\right)\left(\begin{array}{c}
        aA^{(h,l)}\\
        a+B^{(h,l)}\\
        A^{(h,l)}\\
        1+aB^{(h,l)}
    \end{array}\right)
\end{equation}
\begin{equation}
    A^{(h,l)}=\mathcal{B},\quad B^{(h,l)}=\mathcal{A}^{*}.
\end{equation}
\subsubsection{Electron source from right}
For electron source from right, the wavefunction fulfills
\begin{equation}
    \left(\begin{array}{c}
        C^{(e,r)}\\
        1+aD^{(e,r)}\\
        aC^{(e,r)}\\
        a+D^{(e,r)}
    \end{array}\right)=\left(\begin{array}{cccc}
        e^{-i\phi/2}\cosh R & -e^{-i\phi/2}i\sinh R & 0 & 0\\
        e^{-i\phi/2}i\sinh R & e^{-i\phi/2}\cosh R & 0 & 0\\
        0 & 0 & e^{i\phi/2}\cosh R & e^{i\phi/2}i\sinh R\\
        0 & 0 & -e^{i\phi/2}i\sinh R & e^{i\phi/2}\cosh R
    \end{array}\right)\left(\begin{array}{c}
        aA^{(e,r)}\\
        B^{(e,r)}\\
        A^{(e,r)}\\
        aB^{(e,r)}
    \end{array}\right)
\end{equation}
\begin{gather}
    A^{(e,r)}=\mathcal{C}=-\frac{iE_{M}\sqrt{E^{2}-\Delta^{2}}\tanh R}{E^{2}-E_{M}^{2}} \nonumber \\
    B^{(e,r)}=\mathcal{D}^{*}=\frac{(E^{2}-\Delta^{2})E_{M}+iE\sqrt{E^{2}-\Delta^{2}}\Delta\sqrt{D}\sin\frac{\phi}{2}}{\Delta(E^{2}-E_{M}^{2})}.
\end{gather}
\subsubsection{Hole source from right}
For hole source from right, the wavefunction fulfills
\begin{equation}
    \left(\begin{array}{c}
        a+C^{(h,r)}\\
        aD^{(h,r)}\\
        1+aC^{(h,r)}\\
        D^{(h,r)}
    \end{array}\right)=\left(\begin{array}{cccc}
        e^{-i\phi/2}\cosh R & -e^{-i\phi/2}i\sinh R & 0 & 0\\
        e^{-i\phi/2}i\sinh R & e^{-i\phi/2}\cosh R & 0 & 0\\
        0 & 0 & e^{i\phi/2}\cosh R & e^{i\phi/2}i\sinh R\\
        0 & 0 & -e^{i\phi/2}i\sinh R & e^{i\phi/2}\cosh R
    \end{array}\right)\left(\begin{array}{c}
        aA^{(h,r)}\\
        B^{(h,r)}\\
        A^{(h,r)}\\
        aB^{(h,r)}
    \end{array}\right)
\end{equation}
\begin{equation}
    A^{(h,r)}=\mathcal{D},\quad B^{(h,r)}=\mathcal{C}.
\end{equation}
\section{Derivation of Josephson current\label{app.JC}}
In this appendix, we derive the Josephson current formula in Eq.~(\ref{eq.current1}) in Sec.~\ref{sec:bound}. 
By using Eq.~(\ref{eq.bogoliubov}), we can write the field operators for left/right moving electrons in terms Bogoliubov quasiparticle operators,
in terms of coefficients of wave functions derived in the previous section,  
\begin{align}
    \psi_{+}(0^{-})&=\int dE\,
    J(E)\left[(1+a\mathcal{A})\gamma_{(E,e,l)}+a\mathcal{B}\gamma_{(E,h,l)}+a\mathcal{C}\gamma_{(E,e,r)}+a\mathcal{D}\gamma_{(E,h,r)}\right. \nonumber \\
    &-\left.a\mathcal{B}^{*}\gamma_{(E,e,l)}^{\dagger}-(1+a\mathcal{A})\gamma_{(E,h,l)}^{\dagger}-a\mathcal{D}\gamma_{(E,e,r)}^{\dagger}-a\mathcal{C}^{*}\gamma_{(E,h,r)}^{\dagger}\right]+a(E_{M})A_{0}\gamma_{0}-a(E_{M})^{*}B_{0}^{*}\gamma_{0}^{\dagger}            
\end{align}
\begin{align}
\psi_{-}(0^-)&=\int dE\, J(E)\left[\mathcal{B}\gamma_{(E,e,l)}+(a+\mathcal{A}^{*})\gamma_{(E,h,l)}+\mathcal{D}^{*}\gamma_{(E,e,r)}+\mathcal{C}\gamma_{(E,h,r)}\right. \nonumber \\
&+\left.(a+\mathcal{A}{}^{*})\gamma_{(E,e,l)}^{\dagger}+\mathcal{B}^{*}\gamma_{(E,h,l)}^{\dagger}+\mathcal{C}{}^{*}\gamma_{(E,e,r)}^{\dagger}+\mathcal{D}^{*}\gamma_{(E,h,r)}^{\dagger}\right]+B_{0}\gamma_{0}+A_{0}^{*}\gamma_{0}^{\dagger}.
\end{align}
 
At zero temperature, all continuum states with negative eigenvalues of the Bogoliubov-de Gennes Hamiltonian are occupied, 
and all continuum states with positive eigenvalues are empty. The occupation of the Andreev bound state is $n=0,1$, depend
on the fermion parity of the system. 
This leads to the Josephson current, by using Eq.~(\ref{eq.current}), 
\begin{align}
    \braket{I}&=ev_{F}\left[\abs{A_{0}}^{2}-\abs{B_{0}^{2}}\right](2n-1)\nonumber \\
    &=\frac{e\Delta}{2}\sqrt{D}\sin\frac{\phi}{2}(1-2n) \nonumber \\
    &=\frac{\pi G}{2}\frac{\Delta^2\sin\phi}{2eE_M(\phi)}(1-2n), \qquad G=e^2D/\pi. 
\end{align}
To obtain the above equation, we have used the identity
\begin{equation}
    \abs{\mathcal{A}}^{2}+\abs{\mathcal{B}}^{2}+\abs{\mathcal{C}}^{2}+\abs{\mathcal{D}}^{2}=1.
\end{equation}

\section{Derivation of ${\rm Re}Y(\omega)$\label{app.resp}}
In this appendix, we apply linear response theory to derive the real part of the admittance via the response function given in 
Eq.~(\ref{eq.chi}) of Sec.~\ref{sec:admittance}. 
The response function $\chi{(t)}$ can be written as
\begin{align}
    \chi(t) =-i\theta(t)ev_{F}^{2}\left\{ \left\langle [\psi_{+}^{\dagger}(t)\psi_{+}(t),\psi_{+}^{\dagger}\psi_{+}]\right\rangle _{0}
    +\left\langle [\psi_{-}^{\dagger}(t)\psi_{-}(t),\psi_{-}^{\dagger}\psi_{-}]\right\rangle _{0}\right\}  \nonumber \\
    +i\theta(t)ev_{F}^{2}\left\{ \left\langle [\psi_{+}^{\dagger}(t)\psi_{+}(t),\psi_{-}^{\dagger}\psi_{-}]\right\rangle _{0}
    +\left\langle [\psi_{-}^{\dagger}(t)\psi_{-}(t),\psi_{+}^{\dagger}\psi_{+}]\right\rangle _{0}\right\}. 
\end{align}
As a function of Matsubara frequency, it can be written as
\begin{equation}
    \chi(i\Omega_{n})=ev_{F}^{2}\left[\mathcal{G}_{1}(i\Omega_{n})+\mathcal{G}_{2}(i\Omega_{n})-\mathcal{G}_{3}(i\Omega_{n})-\mathcal{G}_{4}(i\Omega_{n})\right] \label{eq.chi_M}
\end{equation}
and the frequency-dependent response function follows from it by analytical continuation. 

\subsection{$\mathcal{G}_{1}(i\Omega_n)$}

For $\tau \geq 0$, 
\begin{equation}
    \mathcal{G}_{1}(\tau)=-\left\langle \mathcal{T}\psi_{+}^{\dagger}(\tau)\psi_{+}(\tau)\psi_{+}^{\dagger}\psi_{+}\right\rangle _{0}
    =-\left\langle \psi_{+}^{\dagger}(\tau)\psi_{+}\right\rangle _{0}\left\langle \psi_{+}(\tau)\psi_{+}^{\dagger}\right\rangle _{0}
    +\left\langle \psi_{+}^{\dagger}(\tau)\psi_{+}^{\dagger}\right\rangle _{0}\left\langle \psi_{+}(\tau)\psi_{+}\right\rangle _{0}.
\end{equation}
By using the relation between electron operators and Bogoliubov quasiparticle operators in  Eq.\ (\ref{eq.bogoliubov}), 
at $T=0$, we have 
\begin{equation}
    \left\langle \psi_{+}^{\dagger}(\tau)\psi_{+}\right\rangle _{0}
    =\int dE P(E)e^{-E\tau}+\abs{B_{0}}^{2}e^{-E_{M}\tau}(1-n)+\abs{A_{0}}^{2}e^{E_{M}\tau}n,
\end{equation}
\begin{equation}
    \left\langle \psi_{+}(\tau)\psi_{+}^{\dagger}\right\rangle _{0}
    =\int dE P(E)e^{-E\tau}+\abs{A_{0}}^{2}e^{-E_{M}\tau}(1-n)+\abs{B_{0}}^{2}e^{E_{M}\tau}n,
\end{equation}
\begin{equation}
    \left\langle \psi_{+}^{\dagger}(\tau)\psi_{+}^{\dagger}\right\rangle _{0}=-B_{0}A_{0}^{*}\left[e^{-E_{M}\tau}(1-n)+e^{E_{M}\tau}n\right],
\end{equation}
\begin{equation}
    \left\langle \psi_{+}(\tau)\psi_{+}\right\rangle _{0}=-A_{0}B_{0}^{*}\left[e^{-E_{M}\tau}(1-n)+e^{E_{M}\tau}n\right],
\end{equation}
with 
\begin{equation}
    P(E)=\frac{1}{2\pi v_{F}}\frac{E\sqrt{E^{2}-\Delta^{2}}}{E^{2}-E_{M}^{2}}.
\end{equation}
Hence, 
\begin{align}
    \mathcal{G}_{1}(\tau)&=-\left\langle \mathcal{T}\psi_{+}^{\dagger}(\tau)\psi_{+}(\tau)\psi_{+}^{\dagger}\psi_{+}\right\rangle _{0} \nonumber \\
        &=-\int dE_{1}dE_{2}\,P(E_{1})P(E_{2})e^{-(E_{1}+E_{2})\tau}
        -(\abs{A_{0}}^{2}+\abs{B_{0}}^{2})\left[\int dE\, (1-n) P(E)e^{-(E+E_{M})\tau}+\int dE\,n P(E)e^{-(E-E_{M})\tau}\right].
\end{align}
Finally, we obtain
\begin{equation}
    \mathcal{G}_{1}(i\Omega_{n})
    =\int dE_{1}dE_{2}\,\frac{P(E_{1})P(E_{2})}{i\Omega_{n}-E_{1}-E_{2}}
    +(\abs{A_{0}}^{2}+\abs{B_{0}}^{2})\int dE\,P(E)\left[\frac{(1-n)}{i\Omega_{n}-E-E_{M}}+\frac{n}{i\Omega_{n}-E+E_{M}}\right].
\end{equation}

\subsection{$\mathcal{G}_{2}(i\Omega_n)$}
For $\tau\geq 0$,
\begin{equation}
    \mathcal{G}_{2}(\tau)=-\left\langle \mathcal{T}\psi_{-}^{\dagger}(\tau)\psi_{-}(\tau)\psi_{-}^{\dagger}\psi_{-}\right\rangle_{0}
    =-\left\langle \psi_{-}^{\dagger}(\tau)\psi_{-}\right\rangle _{0}\left\langle \psi_{-}(\tau)\psi_{-}^{\dagger}\right\rangle_{0}
    +\left\langle \psi_{-}^{\dagger}(\tau)\psi_{-}^{\dagger}\right\rangle _{0}\left\langle \psi_{-}(\tau)\psi_{-}\right\rangle_{0}.
\end{equation}
Consider $T=0$, 
\begin{equation}
    \left\langle \psi_{-}^{\dagger}(\tau)\psi_{-}\right\rangle _{0}
    =\int dE P(E)e^{-E\tau}+\abs{A_{0}}^{2}e^{-E_{M}\tau}(1-n)+\abs{B_{0}}^{2}e^{E_{M}\tau}n,
\end{equation}
\begin{equation}
    \left\langle \psi_{-}(\tau)\psi_{-}^{\dagger}\right\rangle _{0}
    =\int dE P(E)e^{-E\tau}+\abs{B_{0}}^{2}e^{-E_{M}\tau}(1-n)+\abs{A_{0}}^{2}e^{E_{M}\tau}n,
\end{equation}
\begin{equation}
    \left\langle \psi_{-}^{\dagger}(\tau)\psi_{-}^{\dagger}\right\rangle _{0}=A_{0}B_{0}^*\left[e^{-E_{M}\tau}(1-n)+e^{E_{M}\tau}n\right],
\end{equation}
\begin{equation}
    \left\langle \psi_{-}(\tau)\psi_{-}\right\rangle _{0}=A_{0}^*B_{0}\left[e^{-E_{M}\tau}(1-n)+e^{E_{M}\tau}n\right].
\end{equation}
Thus, we have 
\begin{equation}
    \mathcal{G}_{2}(i\Omega_{n})=\mathcal{G}_{1}(i\Omega_{n})=
    \int dE_{1}dE_{2}\,\frac{P(E_{1})P(E_{2})}{i\Omega_{n}-E_{1}-E_{2}}
    +(\abs{A_{0}}^{2}+\abs{B_{0}}^{2})\int dE\,P(E)\left[\frac{(1-n)}{i\Omega_{n}-E-E_{M}}+\frac{n}{i\Omega_{n}-E+E_{M}}\right].
\end{equation}

\subsection{$\mathcal{G}_{3}(i\Omega_n)$}
For $\tau \geq 0$,
\begin{equation}
    \mathcal{G}_{3}(\tau)=-\left\langle \mathcal{T}\psi_{+}^{\dagger}(\tau)\psi_{+}(\tau)\psi_{-}^{\dagger}\psi_{-}\right\rangle_{0}
    =-\left\langle \psi_{+}^{\dagger}(\tau)\psi_{-}\right\rangle _{0}\left\langle \psi_{+}(\tau)\psi_{-}^{\dagger}\right\rangle_{0}
    +\left\langle \psi_{+}^{\dagger}(\tau)\psi_{-}^{\dagger}\right\rangle _{0}\left\langle \psi_{+}(\tau)\psi_{-}\right\rangle _{0}.
\end{equation}
Consider $T=0$, we have
\begin{equation}
    \left\langle \psi_{+}^{\dagger}(\tau)\psi_{-}\right\rangle _{0}= -\int dE\,M(E) e^{-E\tau} -a(E_{M})A_{0}^{*}B_{0}e^{-E_{M}\tau}(1-n)+a(E_{M})^{*}A_{0}^{*}B_{0}e^{E_{M}\tau}n
\end{equation}
\begin{equation}
    \left\langle \psi_{+}(\tau)\psi_{-}^{\dagger}\right\rangle _{0}= \int dE\, M(E) e^{-E\tau} + a(E_{M})A_{0}B_{0}^{*}e^{-E_{M}\tau}(1-n)-a(E_{M})^{*}B_{0}^{*}A_{0}e^{E_{M}\tau}n
\end{equation}
\begin{equation}
    \left\langle \psi_{+}^{\dagger}(\tau)\psi_{-}^{\dagger}\right\rangle _{0}=-\int dE\, Q(E)^* e^{-E\tau}-a(E_{M})\abs{B_{0}}^{2}e^{-E_{M}\tau}(1-n)+a(E_{M})^{*}\abs{A_{0}}^{2}e^{E_{M}\tau}n
\end{equation}
\begin{equation}
    \left\langle \psi_{+}(\tau)\psi_{-}\right\rangle _{0}=\int dE Q(E) e^{-E\tau}+a(E_{M})\abs{A_{0}}^{2}e^{-E_{M}\tau}(1-n)-a(E_{M})^{*}\abs{B_{0}}^{2}e^{E_{M}\tau}n
\end{equation}
where
\begin{gather}
    M(E) = \frac{1}{2\pi v_F} \frac{i E \sqrt{E^2-\Delta^2}\tanh{R}}{E^2-E_M^2}, \nonumber \\
    Q(E)= \frac{1}{2\pi v_F} \frac{E_M\sqrt{E^2-\Delta^2}}{E^2-E_M^2}.
\end{gather}
Hence
\begin{align}
    \mathcal{G}_{3}(\tau)&=\int dE_{1}dE_{2}\, e^{-(E_{1}+E_{2})\tau}\left[M(E_{1})M(E_{2})-Q^{*}(E_{1})Q(E_{2})\right] \nonumber \\
    &+(1-n)a(E_{M})\int dE\,
    e^{-(E+E_{M})\tau}\left[M(E)(A_{0}^{*}B_{0}+A_{0}B_{0}^{*})-Q(E)\abs{B_{0}}^{2}-Q(E)^{*}\abs{A_{0}}^{2}\right] \nonumber \\
    &+na(E_{M})^{*}\int dE\,
    e^{-(E-E_{M})\tau}\left[Q(E)\abs{A_{0}}^{2}+Q(E)^{*}\abs{B_{0}}^{2}-M(E)(A_{0}^{*}B_{0}+A_{0}B_{0}^{*})\right],
\end{align}
\begin{align}
    \mathcal{G}_{3}(i\Omega_{n})&=\int dE_{1}dE_{2}\,\frac{Q^{*}(E_{1})Q(E_{2})-M(E_{1})M(E_{2})}{i\Omega_{n}-E_{1}-E_{2}} \nonumber \\
    &+(1-n)a(E_{M})\int
    dE\,\frac{Q(E)\abs{B_{0}}^{2}+Q(E)^{*}\abs{A_{0}}^{2}-M(E)(A_{0}^{*}B_{0}+A_{0}B_{0}^{*})}{i\Omega_{n}-E-E_{M}} \nonumber \\ 
    &+na(E_{M})^{*}\int
    dE\,\frac{M(E)(A_{0}^{*}B_{0}+A_{0}B_{0}^{*})-Q(E)\abs{A_{0}}^{2}-Q(E)^{*}\abs{B_{0}}^{2}}{i\Omega_{n}-E+E_{M}}.
\end{align}
\subsection{$\mathcal{G}_{4}(i\Omega_n)$}
For $\tau \geq 0$,
\begin{equation}
    \mathcal{G}_{4}(\tau)=-\left\langle \mathcal{T}\psi_{-}^{\dagger}(\tau)\psi_{-}(\tau)\psi_{+}^{\dagger}\psi_{+}\right\rangle _{0}
        =-\left\langle \psi_{-}^{\dagger}(\tau)\psi_{+}\right\rangle _{0}\left\langle \psi_{-}(\tau)\psi_{+}^{\dagger}\right\rangle_{0}
        +\left\langle \psi_{-}^{\dagger}(\tau)\psi_{+}^{\dagger}\right\rangle _{0}\left\langle \psi_{-}(\tau)\psi_{+}\right\rangle_{0}.
\end{equation}
By using the zero-temperature averages of electron operators
\begin{equation}
    \left\langle \psi_{-}^{\dagger}(\tau)\psi_{+}\right\rangle _{0}=\int dE\, M(E)e^{-E\tau} -a^{*}(E_{M})A_{0}B_{0}^{*}e^{-E_{M}\tau}(1-n)+a(E_{M})A_{0}B_{0}^{*}e^{E_{M}\tau}n
\end{equation}
\begin{equation}
    \left\langle \psi_{-}(\tau)\psi_{+}^{\dagger}\right\rangle _{0}=-\int dE\, M(E)e^{-E\tau} + a^{*}(E_{M})A_{0}^{*}B_{0}e^{-E_{M}\tau}(1-n)-a(E_{M})A_{0}^{*}B_{0}e^{E_{M}\tau}n
\end{equation}
\begin{equation}
    \left\langle \psi_{-}^{\dagger}(\tau)\psi_{+}^{\dagger}\right\rangle _{0}
    =\int dE\, Q(E)^* e^{-E\tau}+a^{*}(E_{M})\abs{A_{0}}^{2}e^{-E_{M}\tau}(1-n)-a(E_{M})\abs{B_{0}}^{2}e^{E_{M}\tau}n
\end{equation}
\begin{equation}
    \left\langle \psi_{-}(\tau)\psi_{+}\right\rangle _{0}=-\int dE\, Q(E)e^{-E\tau}-a^{*}(E_{M})\abs{B_{0}}^{2}e^{-E_{M}\tau}(1-n)+a(E_{M})\abs{A_{0}}^{2}e^{E_{M}\tau}n.
\end{equation}
we obtain
\begin{align}
    \mathcal{G}_{4}(\tau)&=\int dE_{1}dE_{2}\, e^{-(E_{1}+E_{2})\tau}\left[M(E_{1})M(E_{2})-Q^{*}(E_{1})Q(E_{2})\right]\nonumber \\
    &-(1-n)a(E_{M})^{*}\int dE\,
    e^{-(E+E_{M})\tau}\left[M(E)(A_{0}^{*}B_{0}+A_{0}B_{0}^{*})+Q(E)\abs{A_{0}}^{2}+Q(E)^{*}\abs{B_{0}}^{2}\right]\nonumber \\ 
    &+na(E_{M})\int dE\,
    e^{-(E-E_{M})\tau}\left[Q(E)^{*}\abs{A_{0}}^{2}+Q(E)\abs{B_{0}}^{2}+M(E)(A_{0}^{*}B_{0}+A_{0}B_{0}^{*})\right],
\end{align}
\begin{align}
    \mathcal{G}_{4}(i\Omega_{n})&=\int dE_{1}dE_{2}\,\frac{Q^{*}(E_{1})Q(E_{2})-M(E_{1})M(E_{2})}{i\Omega_{n}-E_{1}-E_{2}}\nonumber \\ 
    &+(1-n)a(E_{M})^*\int
    dE\,\frac{M(E)(A_{0}^{*}B_{0}+A_{0}B_{0}^{*})+Q(E)\abs{A_{0}}^{2}+Q(E)^{*}\abs{B_{0}}^{2}}{i\Omega_{n}-E-E_{M}} \nonumber \\ 
    &-na(E_{M})\int
    dE\,\frac{Q(E)^{*}\abs{A_{0}}^{2}+Q(E)\abs{B_{0}}^{2}+M(E)(A_{0}^{*}B_{0}+A_{0}B_{0}^{*})}{i\Omega_{n}-E+E_{M}} .
\end{align}

\subsection{${\rm Re}Y(\omega)$}
Plug the expressions for $\mathcal{G}_1$, $\mathcal{G}_2$, $\mathcal{G}_3$ and $\mathcal{G}_4$ into Eq.\ (\ref{eq.chi_M}), and make analytical
continuation $i\Omega_n \rightarrow \omega + i\eta$, where $\eta \rightarrow 0^+$, we obtain the retarded response function
\begin{align}
    \chi(\omega+i\eta) = \chi_{1}(\omega+i\eta)+(1-n)\chi_{2}(\omega+i\eta)+n\chi_{3}(\omega+i\eta).
\end{align}
with
\begin{align}
    \chi_{1}(\omega+i\eta)&=2ev_{F}^{2}\int dE_{1}dE_{2}\,\frac{P(E_{1})P(E_{2})+M(E_{1})M(E_{2})-Q^{*}(E_{1})Q(E_{2})}{\omega+i\eta-E_{1}-E_{2}}\nonumber \\
    &=\frac{eD}{2\pi^{2}}\int_{\Delta}^{\infty}dE_{1}dE_{2}\,\frac{(E_{1}E_{2}-E_{M})\sqrt{E_{1}^{2}-\Delta^{2}}\sqrt{E_{2}^{2}-\Delta^{2}}}{(\omega+i\eta-E_{1}-E_{2})(E_{1}^{2}-E_{M}^{2})(E_{2}^{2}-E_{M}^{2})}
\end{align}
\begin{align}
    \chi_{2}(\omega+i\eta)&=2ev_{F}^{2}\int dE\,\frac{(\abs{A_{0}}^{2}+\abs{B_{0}}^{2})P(E)+{\rm
    Re}\left\{a(E_{M})\left[M(E)(A_{0}^{*}B_{0}+A_{0}B_{0}^{*})-Q(E)\abs{B_{0}}^{2}-Q(E)^{*}\abs{A_{0}}^{2}\right]\right\}}{\omega+i\eta-E-E_{M}}\nonumber \\
    &=\frac{eD}{\pi}\int_{\Delta}^{\infty}dE\,\frac{\sqrt{E^{2}-\Delta^{2}}\sqrt{\Delta^{2}-E_{M}^{2}}}{(\omega+i\eta-E-E_{M})(E+E_{M})}
\end{align}
\begin{align}
    \chi_{3}(\omega+i\eta)&=2ev_{F}^{2}\int dE\,\frac{(\abs{A_{0}}^{2}+\abs{B_{0}}^{2})P(E)+{\rm Re}\left\{
        a(E_{M})\left[Q(E)\abs{B_{0}}^{2}+Q(E)^{*}\abs{A_{0}}^{2}+M(E)(A_{0}^{*}B_{0}+A_{0}B_{0}^{*})\right]\right\}
    }{\omega+i\eta-E+E_{M}}\nonumber \\
    &=\frac{eD}{\pi}\int_{\Delta}^{\infty}dE\,\frac{\sqrt{E^{2}-\Delta^{2}}\sqrt{\Delta^{2}-E_{M}^{2}}}{(\omega+i\eta-E+E_{M})(E-E_{M})}.
\end{align}
To obtain the above expressions, we have used
\begin{equation}
    2{\rm Re}\left[M(E)(A_{0}^{*}B_{0}+A_{0}B_{0}^{*})\right]=\frac{-E\sqrt{E^{2}-\Delta^{2}}\sqrt{\Delta^{2}-E_{M}^{2}}(1-D)}{2\pi v_{F}^{2}\Delta(E^{2}-E_{M}^{2})}
\end{equation}
\begin{equation}
    2{\rm Re}[Q(E)\abs{B_{0}}^{2}+Q(E)^{*}\abs{A_{0}}^{2}]=Q(E)\abs{B_{0}}^{2}+Q(E)^{*}\abs{A_{0}}^{2}=\frac{E_{M}\sqrt{E^{2}-\Delta^{2}}\sqrt{\Delta^{2}-E_{M}^{2}}D}{2\pi v_{F}^{2}\Delta(E^{2}-E_{M}^{2})}
\end{equation}
\begin{equation}
    2P(E)\left(\abs{A_{0}}^{2}+\abs{B_{0}}^{2}\right)=\frac{E\sqrt{E^{2}-\Delta^{2}}\sqrt{\Delta^{2}-E_{M}^{2}}}{2\pi v_{F}^{2}\Delta(E^{2}-E_{M}^{2})}.
\end{equation}

By using the relation ${\rm Re}Y=-(e/\omega){\rm Im}\chi$, we obtain the real part of the admittance
\begin{subequations}
    \label{eq.ReAd}
\begin{gather}
    {\rm Re}Y_{1}=\frac{e^{2}D}{2\pi\omega}\theta(\omega-2\Delta)\int_{\Delta}^{\omega-\Delta}dE\,\frac{\left[E(\omega-E)-E_M^2\right]\sqrt{E^{2}-\Delta^{2}}\sqrt{(\omega-E)^{2}-\Delta^{2}}}{(E^{2}-E_{M}^{2})\left[(\omega-E)^{2}-E_{M}^{2}\right]}\\
    {\rm Re}Y_{2}=e^{2}D\theta(\omega-E_{M}-\Delta)\frac{\sqrt{(\omega-E_{M})^{2}-\Delta^{2}}\sqrt{\Delta^{2}-E_{M}^{2}}}{\omega^{2}} \\
    {\rm Re}Y_{3}=e^{2}D\theta(\omega+E_{M}-\Delta)\frac{\sqrt{(\omega+E_{M})^{2}-\Delta^{2}}\sqrt{\Delta^{2}-E_{M}^{2}}}{\omega^{2}}.
    \end{gather}
\end{subequations}
\end{widetext}

\section{Zeeman field inside a nontopological junction\label{app.B_inside}}

In this appendix, we provide some technical details underlying the results presented in Sec.\ \ref{sec:B_inside}. 

As long as we can neglect spin-orbit and Zeeman coupling in the superconducting leads (but not in the junction region), the subgap spectrum of a nontopological junction can be obtained from the condition \cite{beenakker91}
\begin{equation}
    \det\left(1-\alpha_A^{2}r_{A}^{*}S_{e}r_{A}S_{h}\right)=0.
    \label{eq:ABS_det}
\end{equation}
Here, Andreev reflection from the superconductors is described by
\begin{equation}
    \alpha_A=\frac{E}{\Delta}-i\frac{\sqrt{\Delta^{2}-E^{2}}}{\Delta},\quad r_{A}=e^{i\phi\rho_{z}/2},
\end{equation}
with $\phi$ the phase difference between the two superconductors and $\rho_{j}$ Pauli matrices in left/right space. The normal section of the junction is characterized by the electron and hole scattering matrices $S_e$ and $S_h$. In the presence of Zeeman and spin-orbit coupling, the electron and hole scattering matrices $S_e$ and $S_h$ are $4\times4$ matrices describing the normal section coupled to normal-metal leads and relating outgoing to ingoing channels, with the four components corresponding to left and right channels of either spin. The hole scattering matrix $S_h$ is related to the electron scattering matrix through
\begin{equation}
    S_{h} = \sigma_{y} \left(S_e\right)^* \sigma_{y},
\label{SeSh}
\end{equation}
which follows by particle-hole symmetry. (This uses the same Nambu basis as in Sec.\ \ref{sec:bound}.)

In the short-junction limit, $S_{e}$ and $S_{h}$ can be evaluated at zero energy. In this limit, spin-orbit coupling
leaves the spin degeneracy of the Andreev levels unchanged \cite{Michelsen08,Chtchelkatchev03hjk,Beri2008}. 

Choosing the spin quantization axis along $\mathbf{B}$, the scattering matrices $S_{e}$ and $S_{h}$ are diagonal in the spin indices, with the diagonal entries labeled by $S_{e}^{\sigma}$ and $S_{h}^{\sigma}$ (with $\sigma = \uparrow,\downarrow$). Then, Eq.\ (\ref{eq:ABS_det}) breaks up into two separate determinant equations for the spin components. 

For a single spin channel with transmission $D_{\sigma}=1-R_{\sigma}$, the scattering matrices can be parametrized as
\begin{equation}
    S_{e}^{\sigma}=e^{i\eta_{\sigma}}\left(\sqrt{R_{\sigma}}\rho_{z}e^{i\rho_{z}\gamma_{\sigma}}+\sqrt{D_{\sigma}}\rho_{x}\right).
\end{equation}
Exploiting unitarity and Eq.\ (\ref{SeSh}), we obtain  
\begin{equation}
    \det\left(S_{e}^{\bar{\sigma}}-\alpha^{2}r_{A}^{*}S_{e}^{\sigma}r_{A}\right)=0
\end{equation}
with $\bar{\sigma}=-\sigma$. Focusing on $\sigma=\uparrow$ and the Nambu spinor $(\psi_{\uparrow},\psi_{\downarrow}^\dagger)^T$, the determinant condition becomes
\begin{equation}
    \cos(2\tilde{\alpha}+\tilde{\eta})=R\cos\tilde{\gamma}+D\cos\phi,
\end{equation}
where $\tilde{\eta}=\eta_{\uparrow}-\eta_{\downarrow}$, $\tilde{\gamma}=\gamma_{\uparrow}-\gamma_{\downarrow}$, $D=\sqrt{D_{\uparrow}D_{\downarrow}}$,
$R=\sqrt{R_{\uparrow}R_{\downarrow}}$, and $\alpha=\exp\left(i\tilde{\alpha}\right)$. This equation was derived in Ref.~\cite{Michelsen08}. The corresponding results for $\sigma=\downarrow$ with Nambu spinor $(\psi_{\downarrow},-\psi_{\uparrow}^\dagger)^T$ follow by particle-hole symmetry. If we denote the subgap eigenstates for spin $\sigma$ by $E^\sigma_n(\phi)$, we have $E^\downarrow_n(\phi)=-E^\uparrow_n(\phi)$.

For spin-independent scattering matrices, one has $\tilde{\eta}=0$ and $R+D=1$, and recovers the Andreev bound state given in Eq.~(\ref{eq:ABS_energy}). When the two spin channels are subject to different scattering potentials, we have $R+D<1$ and the energies of the Andreev bound states can be written as
\begin{equation}
    E_{\pm}(\phi)=\Delta\Sgn\left[\sin\left(\frac{\tilde{\eta}}{2}\pm\chi\right)\right]\cos\left(\frac{\tilde{\eta}}{2}\pm\chi\right),
    \label{eq:ABS_spin_energy}
\end{equation}
where
\begin{equation}
    \chi=\frac{1}{2}\arccos\left(R\cos\tilde{\gamma}+D\cos\phi\right).
\end{equation}

\section{Junction based on proximitized Rashba nanowires\label{app.wire}}

In this appendix, we provide some technical details underlying Sec.\ \ref{rashbaJJ}. Consider a Josephson junction formed by two semiconductor nanowires with Rashba spin-orbit coupling, proximity coupled to $s$-wave superconductors and subject to a Zeeman field $B$. For a phase difference of $\phi$, the corresponding Hamiltonian takes the form \cite{lutchyn10}
\begin{align}
    H &= \left(-\frac{\partial_{x}^2}{2m} + i\alpha\sigma_y \partial_{x}-\mu \right)\tau_z 
    + B \sigma_x + \Delta\theta(x-\frac{L}{2})\tau_x  \nonumber \\
    &+ \Delta\theta(-x-\frac{L}{2}) \left(\cos\phi \tau_x + \sin\phi \tau_y \right).
    \label{eq:wire}
\end{align}
where $\alpha$ denotes the strength of the Rashba spin-orbit coupling, $\mu$ the chemical potential, $m$ the effective mass, $L$ the length of the junction, and $\Delta$ the induced pairing strength. We also introduced the Pauli matrices $\sigma_j$ and $\tau_j$ in spin and Nambu space, respectively. 

The bulk dispersion of the model is
\begin{align}
    E_{\pm}(p)^2&=B^2 + \Delta^2 + \xi_p^2  + (\alpha p )^2 \nonumber \\
    &\pm 2\sqrt{B^2\Delta^2 + B^2\xi_p^2 + (\alpha p)^2\xi_p^2 },
\end{align}
where $\xi_{p} = \frac{p^2}{2m} - \mu$. For finite $B$ and $\Delta$, gaps open at $p=0$ and $p=\pm k_F$, where 
\begin{equation}
    k_F = \sqrt{2m(m\alpha^2 + \sqrt{m^2\alpha^4 + B^2})}.
\end{equation}
The gap 
\begin{equation}
    E_{\rm gap}(p=0) = \abs{B - \sqrt{\Delta^2 + \mu^2}}
\end{equation}
at $p=0$ closes for $B=\sqrt{\Delta^2+\mu^2}$ indicating the topological phase transition, with the topological (nontopological) phase corresponding to $B>\sqrt{\Delta^2+\mu^2}$ ($B<\sqrt{\Delta^2+\mu^2}$). The gap 
\begin{equation}
    E_{\rm gap}(p=\pm k_F) = \sqrt{\Delta^2 +2\xi_{k_F}^2 - 2\sqrt{B^2\Delta^2+\xi_{k_F}^{4}}}
\end{equation}
at $\pm k_F$ remains finite throughout.

\bibliographystyle{apsrev4-1}
%
\end{document}